\documentclass{ieeeaccess}

\usepackage{gensymb}
\usepackage{subcaption}
\usepackage{graphicx}
\usepackage{float}
\usepackage{caption}
\usepackage{tabularx}
\usepackage{cite}
\usepackage{amsmath,amssymb,amsfonts}
\usepackage{textcomp}
\usepackage{algorithm, algpseudocode}

\newcolumntype{Y}{>{\centering\arraybackslash}X}

\title{\LARGE \bf
Neutron-Induced, Single-Event Effects \\on Neuromorphic Event-based Vision Sensor: A First Step and Tools to Space Applications
}

\def\BibTeX{{\rm B\kern-.05em{\sc i\kern-.025em b}\kern-.08em
    T\kern-.1667em\lower.7ex\hbox{E}\kern-.125emX}}

\begin{document}
\history{Date of publication xxxx 00, 0000, date of current version xxxx 00, 0000.}
\doi{10.1109/ACCESS.2017.DOI}

\title{Neutron-Induced, Single-Event Effects \\on Neuromorphic Event-based Vision Sensor: A First Step Towards Space Applications}
\author{\uppercase{Seth Roffe}\authorrefmark{1},
\uppercase{Himanshu Akolkar\authorrefmark{2}, Alan D. George\authorrefmark{1}, Bernab{\'e} Linares-Barranco\authorrefmark{5}, and Ryad Benosman}.\authorrefmark{2, 3, 4}}
\address[1]{University of Pittsburgh, 4420 Bayard St. Suite 560, Pittsburgh, PA 15213 (emails: \{seth.roffe,alan.george\}@pitt.edu)}
\address[2]{University of Pittsburgh, Biomedical Science Tower 3, Fifth Avenue, Pittsburgh, PA 15260  (emails: \{akolkar,benosman\}@pitt.edu) }
\address[3]{INSERM UMRI S 968; Sorbonne Universit\'e, UPMC Univ. Paris 06, UMRS 968; CNRS, UMR 7210, Institut de la Vision, F-75012, Paris, France}
\address[4]{Carnegie Mellon University , Robotics Institute, 5000 Forbes Avenue Pittsburgh PA 15213-3890, USA}
\address[5]{Instituto de Microelectr{\'o}nica de Sevilla, CSIC and Universidad de Sevilla, Sevilla, Spain (email: bernabe@imse-cnm.csic.es)}
\tfootnote{This research was supported by SHREC industry and agency members and by the IUCRC Program of the National Science Foundation  under Grant No. CNS-1738783. This work was performed, in part, at the Los Alamos Neutron Science Center (LANSCE), a NNSA User Facility operated for the U.S. Department of Energy (DOE) by Los Alamos National Laboratory (Contract 89233218CNA000001). }

\markboth
{Author \headeretal: Roffe et. al.}
{Author \headeretal: Preparation of Papers for IEEE TRANSACTIONS and JOURNALS}

\corresp{Corresponding authors: Seth Roffe (e-mail: seth.roffe@pitt.edu), Ryad B. Benosman (e-mail: benosman@pitt.edu).}

\begin{abstract}

In this paper, we study the suitability of neuromorphic event-based vision cameras for spaceflight, and the effects of neutron radiation on their performance. Neuromorphic event-based vision cameras are novel sensors that implement asynchronous, clockless data acquisition, providing information about the change in illuminance greater than $(\ge120dB)$ with sub-millisecond temporal precision. These sensors have huge potential for space applications as they provide an extremely sparse representation of visual dynamics while removing redundant information, thereby conforming to low-resource requirements. An event-based sensor was irradiated under wide-spectrum neutrons at Los Alamos Neutron Science Center and its effects were classified. Radiation-induced damage of the sensor under wide-spectrum neutrons was tested, as was the radiative effect on the signal-to-noise ratio of the output at different angles of incidence from the beam source. We found that the sensor had very fast recovery during radiation, showing high correlation of noise event bursts with respect to source macro-pulses. No statistically significant differences were observed between the number of events induced at different angles of incidence but significant differences were found in the spatial structure of noise events at different angles. The results show that event-based cameras are capable of functioning in a space-like, radiative environment with a signal-to-noise ratio of 3.355. They also show that radiation-induced noise does not affect event-level computation. Finally, we introduce the Event-based Radiation-Induced Noise Simulation Environment (Event-RINSE), a simulation environment based on the noise-modelling we conducted and capable of injecting the effects of radiation-induced noise from the collected data to any stream of events in order to ensure that developed code can operate in a radiative environment. To the best of our knowledge, this is the first time such analysis of neutron-induced noise analysis has been performed on a neuromorphic vision sensor, and this study shows the advantage of using such sensors for space applications.
\end{abstract}
\maketitle
\thispagestyle{empty}



\section{INTRODUCTION}
Neuromorphic event-based cameras are remarkably efficient, robust, and capable of operating over a large range
of light intensities. These sensors replicate the design of biological retinas to make full use of their power efficiencies, sparse output, large dynamic range, real-time computation, and low-data bandwidth. Neuromorphic sensors are built by copying aspects of their biological counterparts, and are therefore massively parallel and highly non-redundant \cite{Posch14}. Each pixel of the sensor works independently, sensing changes in light and providing output in the form of discrete events signifying increasing or decreasing light intensity.\par
Event-based cameras are a perfectly suited to space missions where the resource budget is limited and radiation can have catastrophic effects on hardware. These sensors have the potential to improve numerous space applications, including those involved in space situational awareness, target tracking, observation and astronomical data collection \cite{Oltrogge19}. Due to the harsh conditions entailed, however, the performance of such sensors in space is yet to be explored. The scope of this work is to test the resilience of neuromorphic sensors to neutrons impacting the sensor in a highly radiative environment. The goal is to determine the failure modes of the neuromorphic camera as seen under the same spectrum as that produced by cosmic rays and to measure the possible impact of neutrons on the temporal precision of output events, noise levels, and computation.\par
Although studies have been carried out into the behavior of various optoelectronic devices under neutron radiation \cite{Watts97}\cite{El_Mashade04}\cite{Kovalyuk05}\cite{Vujisic10}\cite{Alexander03}\cite{Iricanin99}, no work to date has addressed the radiation-tolerance aspects of event-based visual sensors to analyze if this technology is capable of retaining its efficacy under radiative conditions. To observe and evaluate single-event effects, we irradiated a neuromorphic event-based sensor at Los Alamos National Lab's (LANL) ICE-II neutron facility.\par
The measured neutron energy distribution at LANL-ICE-II is significantly more intense than the flux of cosmic-ray-induced neutrons, and this allows for testing at greatly accelerated rates. An ICE-II radiation test of less than an hour is equivalent to many years of neutron exposure due to cosmic-rays \cite{Nowicki17}. Neutrons are known to interact with the materials in the semiconductor and produce daughter particles, which may deposit or remove charge in sensitive volumes of the chip. If the deposited charge is significant enough, it can change the state of a bit in the system. In a digital system this change of state is known as a bit-flip. Sensors include analog circuitry, and therefore produce more complex behavior than simple bit-flips. Beam testing is popular in sensor processing to classify single-event effects (SEEs) in new computing systems and test the robustness of systems to single-event upsets (SEUs). Different systems may respond in different ways to the radiation that brings about SEEs, producing faults and errors of varying degrees. The affect of SEEs can range from negligible, where an unused area of memory is affected, to single-event latch-ups that could damage the system permanently.\par

Knowing how a system may respond to radiation is vital to the success of a space mission insofar that it provides an overview of the kind of upsets that may arise. This information allows designers to plan for any problems that may be encountered in flight. Single-event upsets (SEUs) are transient in that they do not permanently damage the device, but they may cause some silent data or control errors which, if uncaught, may lead to a loss of performance or accuracy. To reduce risk, it is therefore vital to know how a new system will respond to radiation before deployment.\par

In this paper, we measured the effect of radiation and categorized the SEEs observed in the sensor. We also tested how radiation affects  pure event-based computation in the context of optical flow estimation, which is known to be sensitive to noise and temporal imprecision, under both radiation and non-radiation conditions. Finally, we also used this preliminary data to develop a simulator that makes it possible to inject events with radiation-noise effects into any data stream. We call this simulator the "Event-based Radiation-Induced Noise Simulation Environment," (Event-RINSE). Event-RINSE allows realistic neutron beaming effects to be added to any event based data sequence. These simulated radiation effects enable designers to test developed algorithms prior to mission deployment.




\section{Background}

This section gives an overview of the neuromorphic event-driven visual sensor, its data acquisition principles, and its data types. The use of event-driven sensors for space applications is also discussed.

\begin{figure*}[h]
\centering
\includegraphics[width=1.8\columnwidth]{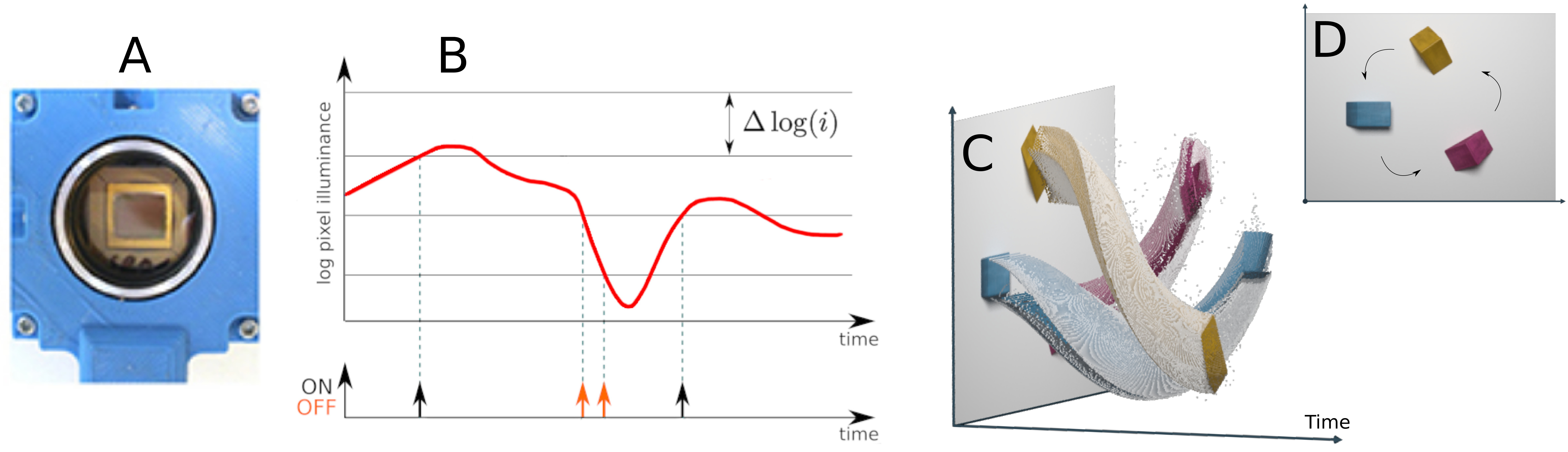}
\caption{Event-based sensor operating principles: (A) The event-based sensor used in this experiment. (B) When a given pixel's luminosity change reaches a given threshold, it produces a visual 
event with an x and y address, a timestamp, and a polarity, which is either ON or OFF depending on the change in relative luminosity. (C,D) The stream of events generated by three rotating shapes, shown here in a color version of the sensor's absolute light measurement output that comes with every event.}
\label{fig:Fig3}
\end{figure*}

\subsection{Neuromorphic Event-Driven Visual Sensors}
Biomimetic, event-based cameras \cite{Lichtsteiner2008} are a novel type of vision sensors that, like their biological counterparts, are made of independent cells/pixels which are driven by events taking place in their field of view, generating an asynchronous stream of spikes/events. This method of data collection is in contrast to conventional vision sensors which are driven by artificially created timing and control signals (frame clock) to create full images that have no relation to either the content or the temporal dynamics of the visual scene. Over the past few years, several types of these event-based cameras have been designed. These include temporal contrast vision sensors sensitive to change in relative luminance, gradient-based sensors sensitive to static edges, devices sensitive to edge-orientation, and optical-flow sensors.

Most of these vision sensors output visual information about the scene in the form of discrete events using Address-Event Representation (AER)~\cite{Mahowald1992}\cite{Lazzaro1995}\cite{Boahen2000}. The data encodes the visual information by sending out tuples $[x;y;t;p]$ --- of space (the pixel where change occurred), time (when the change occurred), and polarity (whether luminance increased or decreased) --- as ON or OFF events, respectively. The event-based camera used in this work is a time-domain encoding event-based sensor with VGA resolution. The sensor contains a 640$\times$480 array of fully autonomous pixels, each relying on an illuminance-change detector circuit. In this study, we will only consider the luminance change circuit that is common to all existing event-based sensors \cite{posch2011qvga}.\par 

The operating principle of an event-based pixel is shown in Figure \ref{fig:Fig3}. The change detector of each pixel individually detects a change in brightness in the field-of-view. Since event-based cameras are not clocked like conventional cameras, the timing of events can be conveyed with a very accurate temporal resolution in the order of microseconds and below\footnote{The highest reported neuromorphic sensor event output rate to date is $1.3\times 10^9$ events per second\cite{samsung2020}.}.\par

These sensors capture information predominantly in the time domain as opposed to conventional frame-based cameras, which currently provide greater amount of spatial information. Since the pixels only detect temporal changes, redundant information like static background is not captured or communicated, resulting in a sparse representation of the scene. Consequently, event-based cameras can have a high temporal-resolution with a very low data-rate \cite{benosman2012asynchronous} compared to conventional cameras, thus conforming to low-resource requirements. Since the pixels are independent of one another and do not need a clock, an error in a few of them will not lead to a catastrophic failure of the device and the sensor will be able to remain operational. \par



\subsection{Conventional Space Situational Awareness}
Space situational awareness (SSA) has been an important topic in military applications for many years \cite{Oltrogge19}\cite{Kennewell13}\cite{Gasparini10}\cite{Pelton19}. SSA is the ability to detect and keep track of surrounding objects and debris to avoid collisions. For SSA, vision systems with high temporal-resolution and low latency are required to accurately detect objects. Event-based cameras are therefore the perfect candidate to replace limited conventional sensing methods in satellite awareness.\par

Ender et al. \cite{ender2011radar} details the use of radar in SSA for collision detection, orbit estimation, and propagation. The benefit of radar is that it has a very large coverage, meaning it can consistently observe a wide area in an arc of almost 5000 km. However, since radio uses long wavelengths, this methodology would only work for larger objects \cite{ender2011radar}. Smaller objects would be impossible to detect via radio waves.\par

One difficulty in object detection to avoid collisions in space is the modeling of non-linear orbits in real-time. Several methods have been proposed to predict non-linear orbits for SSA. One is to use Gaussian mixture modeling to exploit properties of linear systems to extrapolate information about a non-linear system, and then to use Gaussian splitting to reduce the errors induced by that extrapolation \cite{demars2010nonlinear}. The mixture model enables complex, non-linear orbits to be mapped more accurately, providing a better judgment of potential collisions. The issue arises when this kind of surveillance for object avoidance needs to be done autonomously. The calculations presented are too complex to be performed efficiently by a satellite's embedded platform. Also, since the analysis carried out by such platforms is based on statistical manipulation, it needs to be verified by human intervention in order to avoid any statistical anomalies that may cause potential collisions.\par

Abbot and Wallace \cite{abbot2007decision} tackle the SSA problem of decision support for tracking large amounts of orbiting space debris. They claim that the limited number of sensors leads to inconsistent surveillance of the objects under observation, and therefore propose a cooperative monitoring algorithm for geosynchronous earth orbit satellites to address collision prevention and provide automated alerts. However, this methodology relies on Bayesian modeling, which can be computationally intensive for embedded platforms and requires publicly available data to create the models. With satellites of unknown orbits, unexpected collisions could therefore become an issue.\par

These techniques also require fast positional capture of the observed objects which is difficult with the video cameras currently available for space exploration. Event-based cameras could fill this space by providing low latency/resources sensing for SSA.

\subsection{Event based sensors for Space Situational Awareness}

The high dynamic range of event-based sensors with both low-light and bright-light sources allows visual information to be inferred even in the darkness of space or when a bright sun is in the sensor's field-of-view (FoV). It also means that the area around the sun can be observed, even when the sun is coming up over the horizon of a satellite's orbit.\par

The use of event-based cameras in space-related applications is not well developed. Most of the work has been carried out in the context of terrestrial telescope observation of low brightness objects in Low-Earth Orbit (LEO) and Geostationary-Earth Orbit (GEO) \cite{Ralph19}\cite{Afshar20}.\par

Event-based cameras can offer a promising solution to collision avoidance in space provided their high temporal precision and sparsity of data are properly taken into account when designing algorithms. The current trend of generating frames of events, and gray levels to recycle decades of conventional computer vision and machine learning techniques has led to their being used as simple high dynamic range conventional cameras. In this work we focus only on the temporal properties of these sensors, considering cases of per-event computation that preserve the temporal properties of event-based cameras that have been shown to be the key to developing new applications \cite{Akolkar}.\par

There has been extensive research into event-based cameras for real-time tracking and low-power computer systems within the last decade. Many algorithms have been developed that allow for objects to be tracked within the visual space of an event-driven sensor. Reverter et al. developed one such method that makes it possible to track many different shapes, as long as the pattern of the shapes is known a-priori \cite{Reverter2015}. Similarly, Lagorce et al. provide a multi-kernel Gaussian mixture model tracker for the detection and tracking of different shaped objects \cite{Lagorce2015}. Other methods use spatial matching to allow object tracking even in occluded conditions \cite{Ni2015}\cite{camu_stereo} and provide haptic stability by tracking gripper positions in microrobotics applications \cite{Ni2012}. The low computational requirements of event-based sensors even allow tracking systems to be implemented on embedded platforms \cite{Litzenberger2006} and on FPGAs \cite{Barranco2015}. Newer improved spatio-temporal feature detection could improve these methods further \cite{Lagorce2015_feature}. Novel methods can even detect and track objects in conditions where both the camera and the objects are moving independently \cite{Ni2012,Ramesh2018}\cite{Mitrokhin2018}.

\subsection{Neutron-Beam Testing}
Srour and McGarrity \cite{srour1988radiation} detail the effects of space radiation on microelectronic circuits, discussing damage, ionization, and SEEs on optoelectronic devices.
Modern models describe the most of the radiation experienced in the space environment as consisting of protons and heavy ions \cite{badhwar1992improved}. However, this experiment primarily uses wide-spectrum neutrons to test the sensor of interest. In general, neutron beam testing is useful for classifying single-event effects in electronics. Since interest is focused on the response of the device, the source of the upsets become irrelevant. Neutron testing is also useful to test the robustness of systems to SEUs. As an example, NASA Langley Research Center and Honeywell performed neutron beam tests to study the robustness of their flight control computer architecture \cite{eure2004closed}. Their primary goal was to show that they were able to recover from neutron-induced SEUs. The recovery demonstrated system's capabilities in a hazardous environment, even though the radiation source was not neutrons.

When radiation impacts a device, energy is deposited into the target material, causing various faults in the hardware. These faults can have different effects such as memory corruption or glitches in analog and digital hardware  \cite{bagatin2018ionizing}. In an imaging sensor, these errors would manifest as corrupted pixels or improper output. One type of effect, single-event effects (SEEs), occurs when a high-energy particle strikes a microelectronic component and changes a single state of the internals in the device \cite{srour1988radiation}. These effects can then manifest as transient-data errors, corrupting the data output.\par


\section{METHODOLOGY}

This section gives an overview of how the radiation experiment was performed, explaining the Los Alamos Neutron Science Center's neutron beam and detailing how data was collected during irradiation.

\begin{figure}[hbtp]
    \centering
    \includegraphics[width=0.4\textwidth]{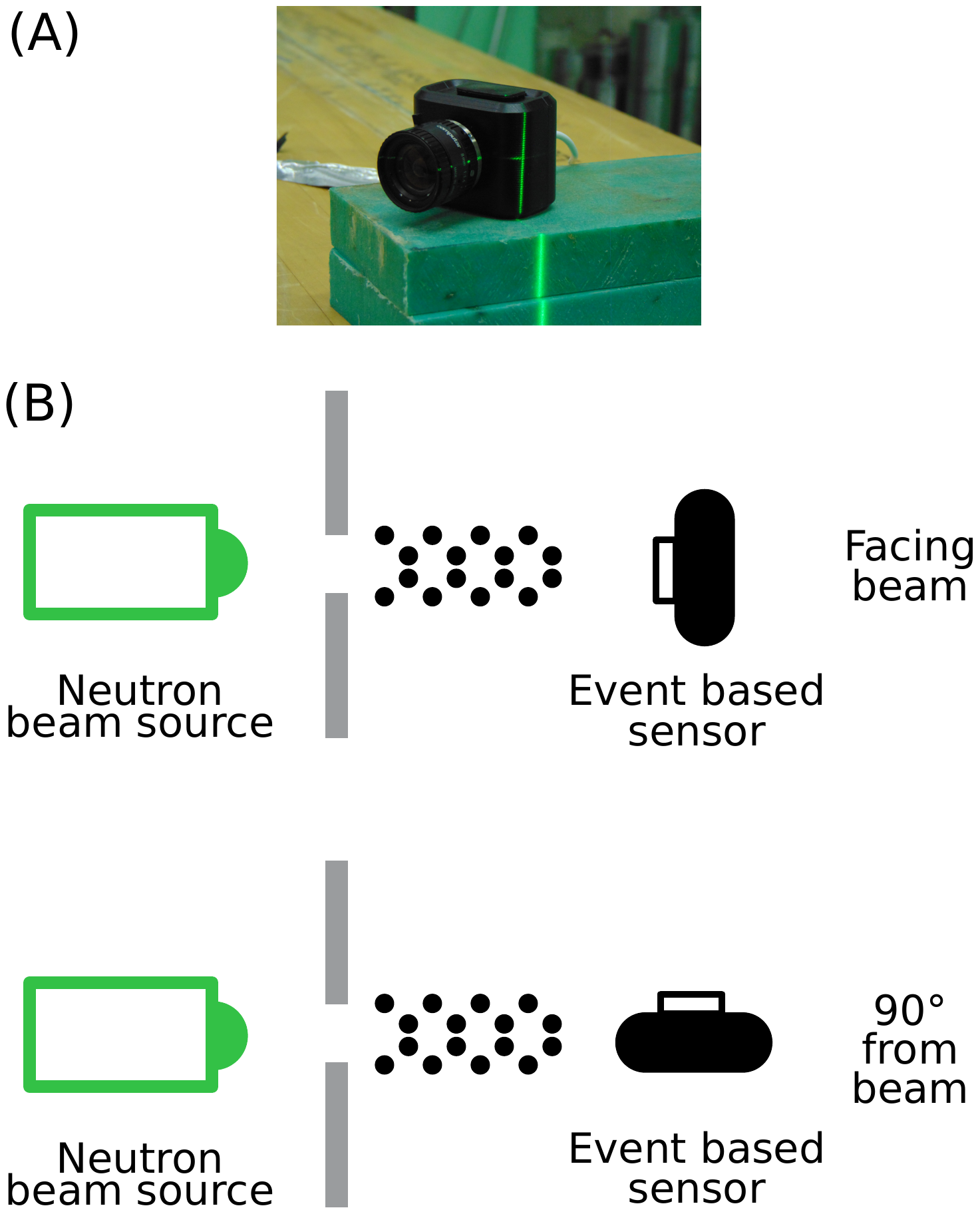}
    \caption{(A) The event-driven sensor under test sitting on a stand that is non-reactive to neutron radiation. To ensure that the neutrons passed through the sensor, the green laser was used to aim the beam. (B) Schematics showing the sensor placed at a fixed distance from the beam source in two conditions - facing the beam directly and at a 90 $\degree$ angle of incidence.}
    \label{fig:sensor}
\end{figure}

\subsection{Event-Camera}
{The sensor used for the experiments in this paper was an event-based sensor based on \cite{posch2011qvga} with VGA resolution (640$\times$480 pixels) fabricated in 180nm CMOS-CIS technology. The chip has a total die size of 9.6$\times$7.2mm$^2$, with a pixel size of 15$\times$15$\mu m^2$, and a fill factor (ratio of photo-diode area over total pixel area) of 25\%. The maximum event-rate for this camera is specified as 66~Meps (mega events per second). During recordings, output events were time-stamped with micro-second resolution by the camera interface and communicated via USB to a host computer for storage. In our recordings we observed a maximum of about 30 events captured with the same micro-second timestamp, meaning that the maximum sensor throughput was not reached.}

\subsection{Irradiation}
The event-camera under test was irradiated at ICE-II, Los Alamos Neutron Science Center's wide-spectrum neutron-beam facility. The Los Alamos Neutron Science Center (LANSCE) provides the scientific community with intense sources of neutrons, which can be used to perform  experiments  supporting  civilian  and  national  security  research.  The ICE facility was built to perform accelerated neutron testing of semiconductor devices. Flight Path 30L and 30R, known as ICE House and ICE-II, allow users to efficiently set up and conduct measurements \cite{Nowicki17}. The sensor was irradiated for two days, from November 23, 2019 to November 24, 2019 under wide-spectrum neutrons of energies ranging from $0.1 MeV$ to $>600 MeV$. The general setup is shown in Figure \ref{fig:sensor}.\par

An event-based camera was placed at a fixed distance in the beam to act as a control on the effective neutron flux. The sensor was placed at different angles of incidence from the beam as shown in Fig. \ref{fig:sensor}(B) to detect any potential differences in the effects observed. Data was collected at an angle of $90^{\degree}$ from the beam and directly facing the beam source. \par

In this experiment, the event-camera was irradiated with the lens cap on to avoid any light or environmental noise on the sensor. Thus, the noise recorded from the sensor in this experiment primarily come from the effects of the radiation rather than those induced by the light sources in the environment.

\begin{figure}[]
    \centering
    \includegraphics[width=.99\columnwidth]{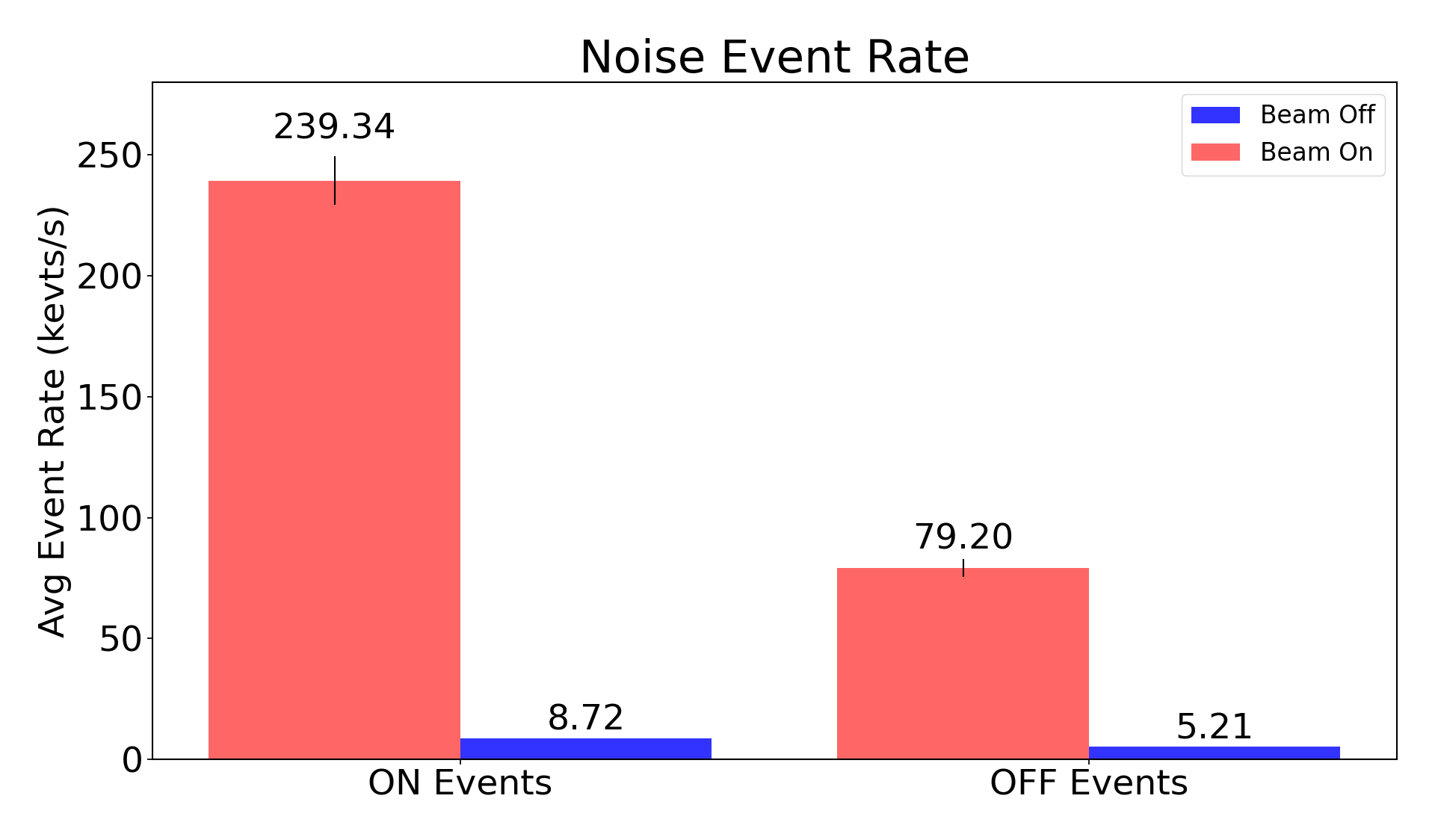}
    \caption{Average number of noise events per second induced due to radiation compared to noise without irradiation over 2 days of irradiation. The recordings were taken with the lens cap on the camera, so the induced events were due either to the inherent thermal noise or to noise induced through the neutrons. Radiation induced more ON events than OFF events (3:1 ratio). }
    \label{fig:totEvtsOn}
\end{figure}

\subsection{Data Collection and analysis}

The sensor was connected to a computer running software which interfaced with the sensor to record events. Events were later processed offline. 
Data was taken with the beam on and off in order to observe the increase in noise caused by irradiation. Radiation-induced noise can be seen in the form of clustered noise-like patterns and line streaks of moving particles in the focal plane, as will be detailed in the following sections. 
The recorded data was parsed to get an event rate to measure the number of events generated by the sensor per second. The counted events were then separated into ON and OFF events. The average events per second were calculated for each experiment with standard deviation as error.

\begin{figure}[h]
    \centering
    \includegraphics[width=0.5\textwidth]{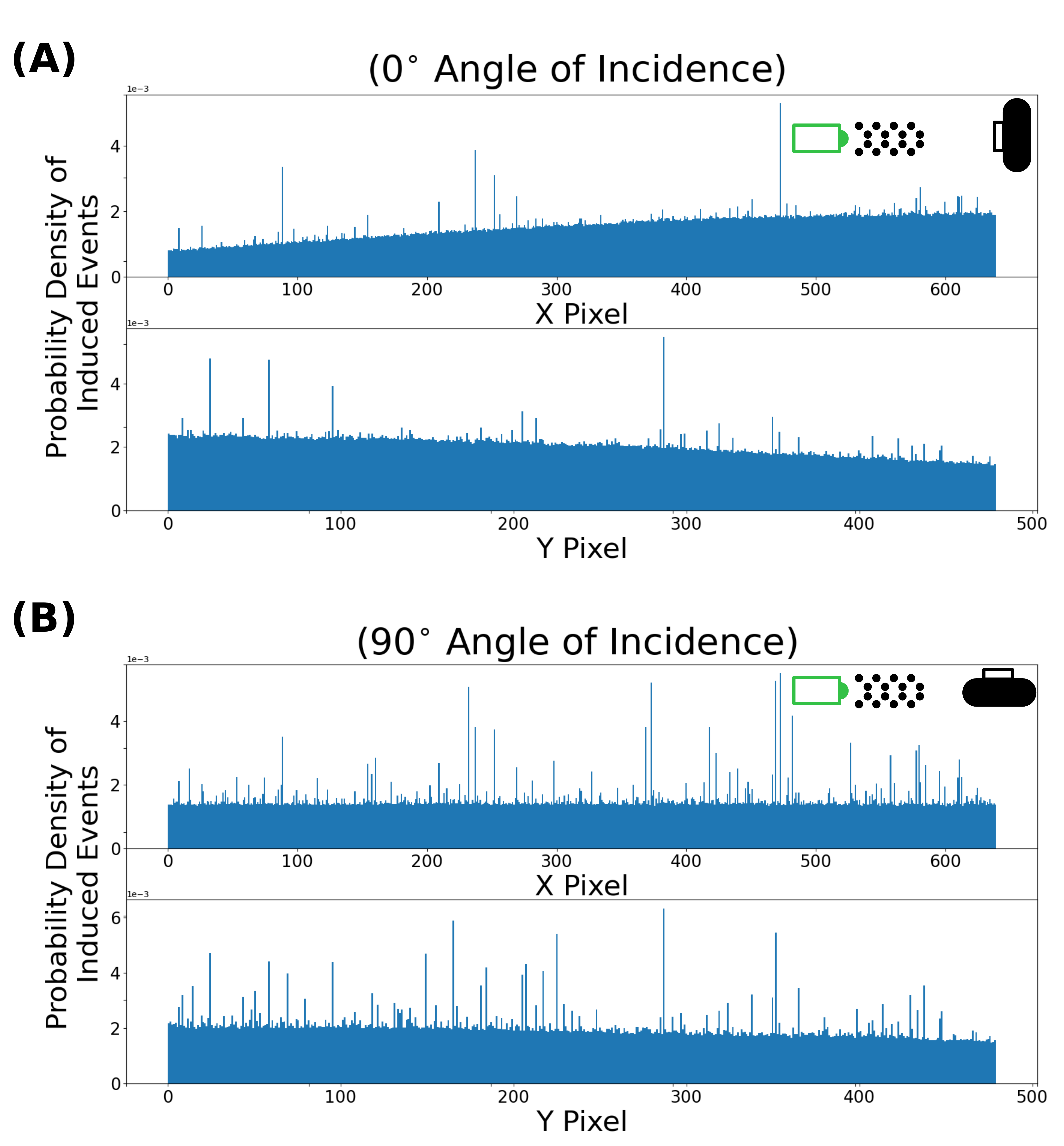}
    \caption{Probability density of events by location on the sensor for (A) 0 degree angle of incidence and (B) 90 degree angle of incidence. The graphs show that the entire sensor was radiated uniformly over the field of view for both conditions. }
    \label{fig:density0deg}
\end{figure}

Data was collected with the sensor facing the beam source and at $90^\degree$, to observe how the angle of incidence affected the incoming radiation noise. The number of events was measured for both ON and OFF events in each orientation and compared. A Mann-Whitney U test was used to determine statistical significance in the differences between the two orientation distributions \cite{Mann47}.

This experiment measured patterns influenced by the effective neutron flux and the number of ON events and OFF events. The patterns were analyzed using an understanding of the sensor's internal circuitry to determine the physical effect of radiation on the sensor. This methodology presents a categorization of SEEs in the form of radiation-induced noise.

To ensure the radiation-induced noise would not overwhelm signal integrity, a pendulum was placed in the visual field to measure the signal-to-noise ratio. Since the signal could be observed with and without radiation-induced noise, the signal-to-noise ratio could be calculated by simply dividing the number of signal events by the noise events produced by radiation. This ratio could then be used to determine the robustness of the sensor to radiation in terms of loss of signal integrity. To validate the signal-to-noise ratio, a correlation test was performed between the radiated data and the non-radiated data. 

\section{RESULTS}

This section gives an overview of the results of the radiation experiment, discussing noise rates, patterns, and analyses.

\subsection{Induced-Event Rate}
Data was collected with the lens cap on the sensor to minimize environmental influence from external lighting. First, the mean number of radiation-induced ON and OFF events per second was measured. The average number of events can be seen in Figure \ref{fig:totEvtsOn}. A significant bias towards ON events was observed.

The induced-event probability density was plotted against the pixel coordinates of the sensor to observe any location preferences for upsets. To measure this, the pixel location of each induced event was divided by the total number of events measured for both angles of incidence. These measurements can be seen in Figures \ref{fig:density0deg}(A) and \ref{fig:density0deg}(B).

In both cases, the induced events are quite uniform across the sensor, with the $0\degree$ angle of incidence tending to bias towards the location of the neutron beam's 1 inch diameter. We can see that about twice as many events were produced for high $x$ and low $y$ values than for the opposite corner. However, this is due to human error in placing the sensor in the beam path. In other words, there is no particular area of the sensor that is more vulnerable to neutron radiation effects than other areas. This is further demonstrated in the $90\degree$ angle of incidence result. Every pixel across the sensor showed a similar response.

\begin{figure}[h]
    \centering
    \includegraphics[width=0.45\textwidth]{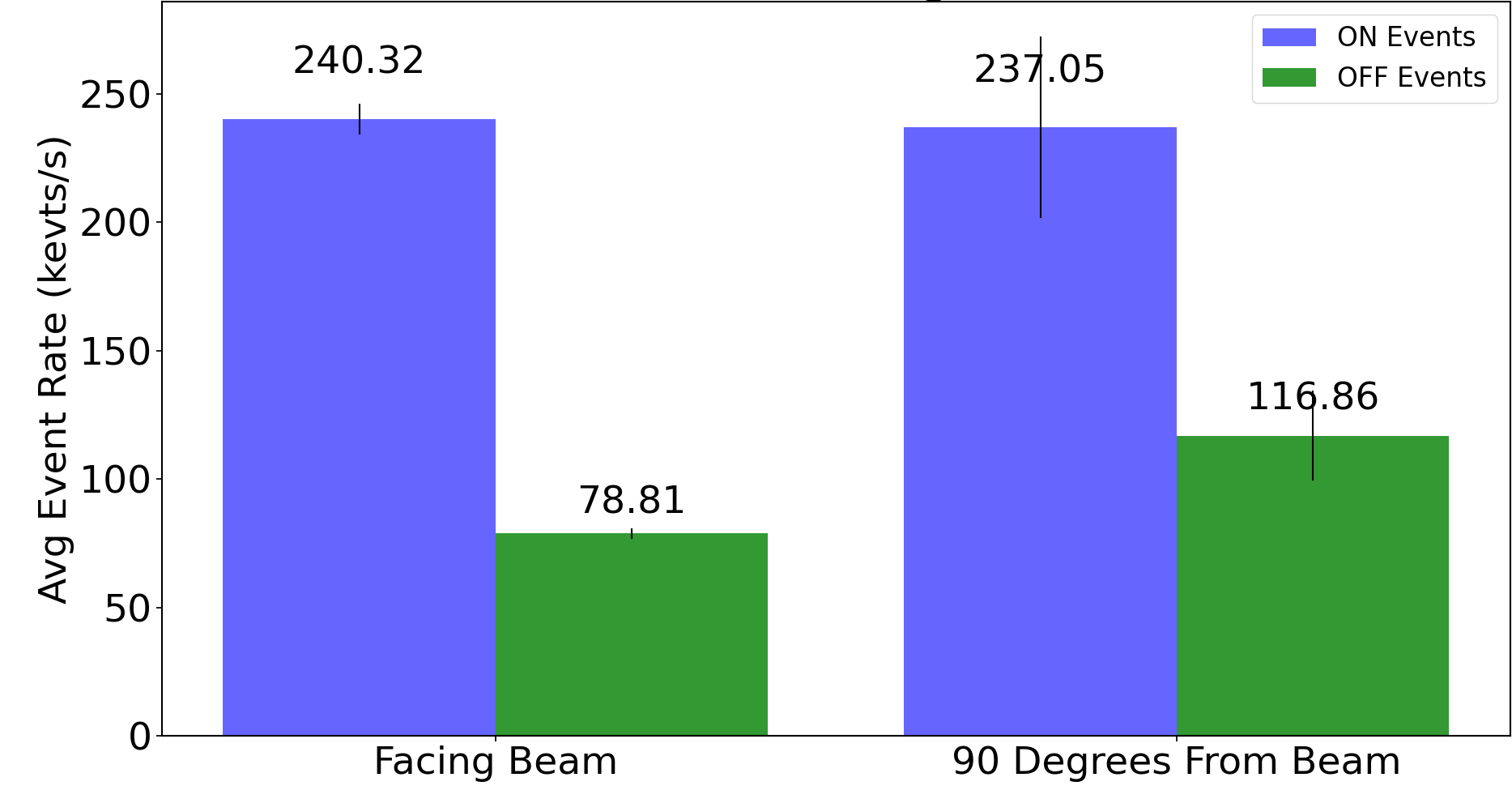}
    \caption{Events observed at different angles of incidence. Data was collected at $90^{\degree}$ from the beam and facing directly towards the beam. No significant difference was found between the number of noise events generated for the two conditions even though the sensor would be expected to interact with more neutrons when facing the beam. At $0\degree$, more events were produced at high $x$ and low $y$ values than for the opposite corner. This is the result of human error in placing the sensor in the beam path. As expected, every pixel across the sensor showed a similar response.
    }
    \label{fig:orientationEvts}
\end{figure}

\subsection{Angle of Incidence Comparison}
Data was collected at two orientations: facing the beam with an angle of incidence of $0^{\degree}$ and at an angle of incidence of $90^{\degree}$ from the beam source. These two distributions were then analyzed separately to observe any significant differences.

Figure \ref{fig:orientationEvts} shows that there was a slight difference between the number of OFF events per second induced between the two orientations. A Mann-Whitney U test was performed on the two distributions to test for statistical significance \cite{Mann47} but no statistically significant difference was found.

\begin{figure}[h]
    \centering
    \includegraphics[width=0.45\textwidth]{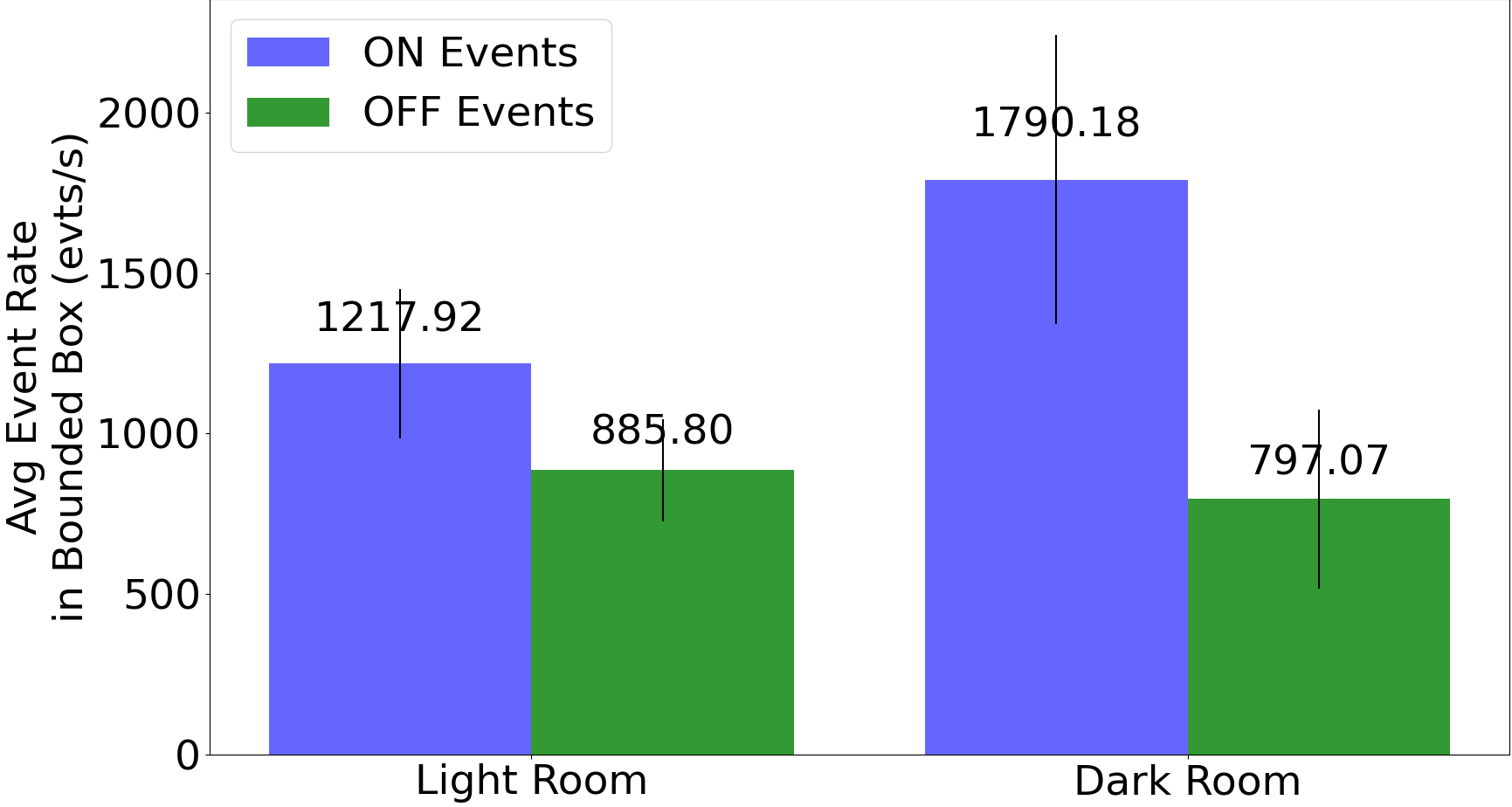}
    \caption{Number of events induced in a $50\times50$ pixel bounded box for a light room vs a dark room. Given the contrast sensitive nature of the sensor, and as  expected, we observed that more ON noise events were generated in the case of dark room since the neutron interactions allowed for the event generation threshold to be crossed more often. The OFF noise events did not increase significantly.}
    \label{fig:lightvsdark}
\end{figure}
\begin{figure}[h]
    \centering
    \includegraphics[width=0.45\textwidth]{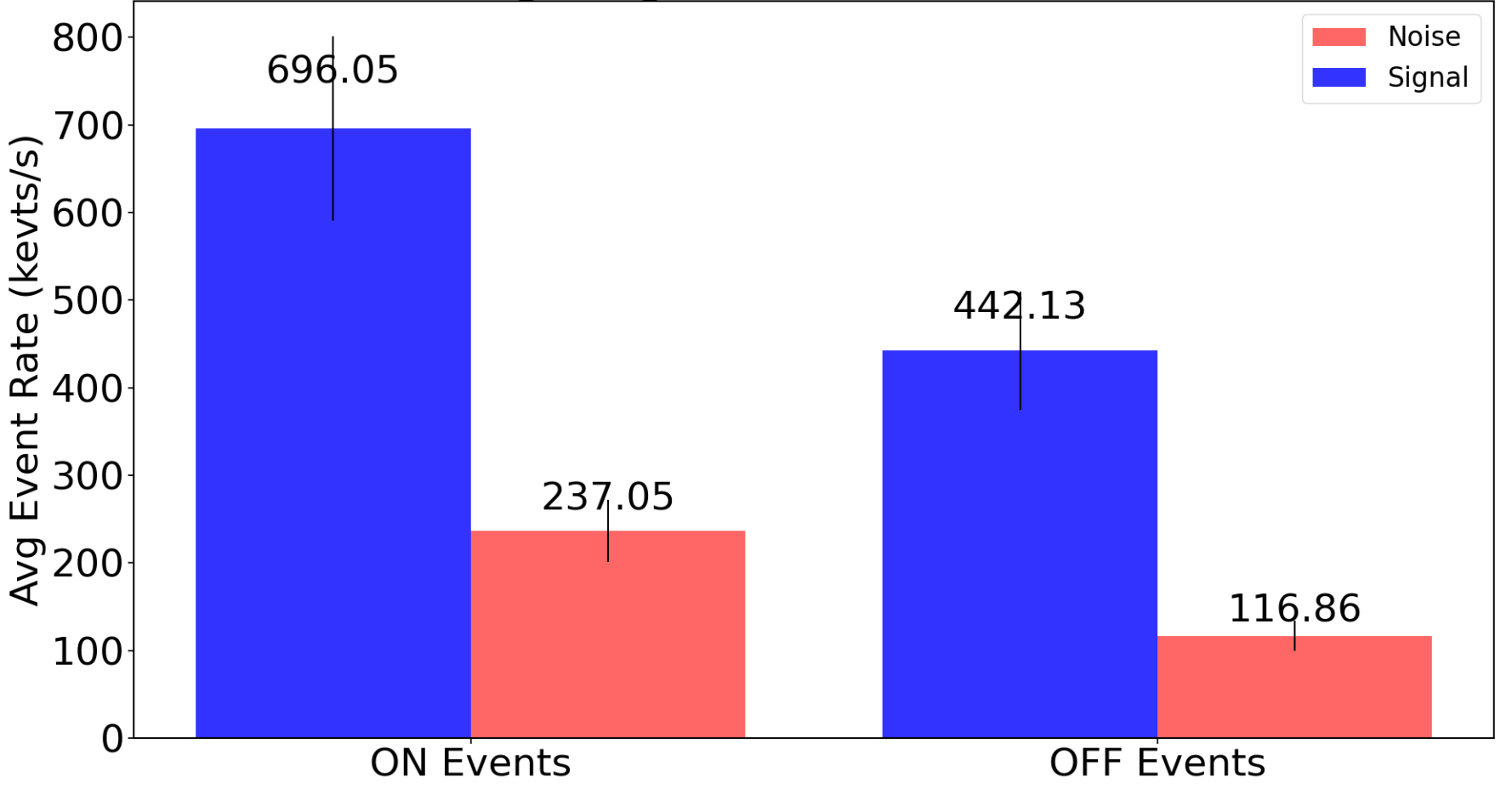}
    \caption{Number of signal events observed vs radiation-induced noise events. Signal events were calculated as the rate of events while recording a cyclic pendulum where as the noise rate was computed from isolated radiation induced events. The signal-to-noise ratio for the sensor even under strong neutron radiation was found to be 3.355.}
    \label{fig:snr}
\end{figure}

\begin{figure*}[h]
\centering
    \includegraphics[width=0.9\textwidth]{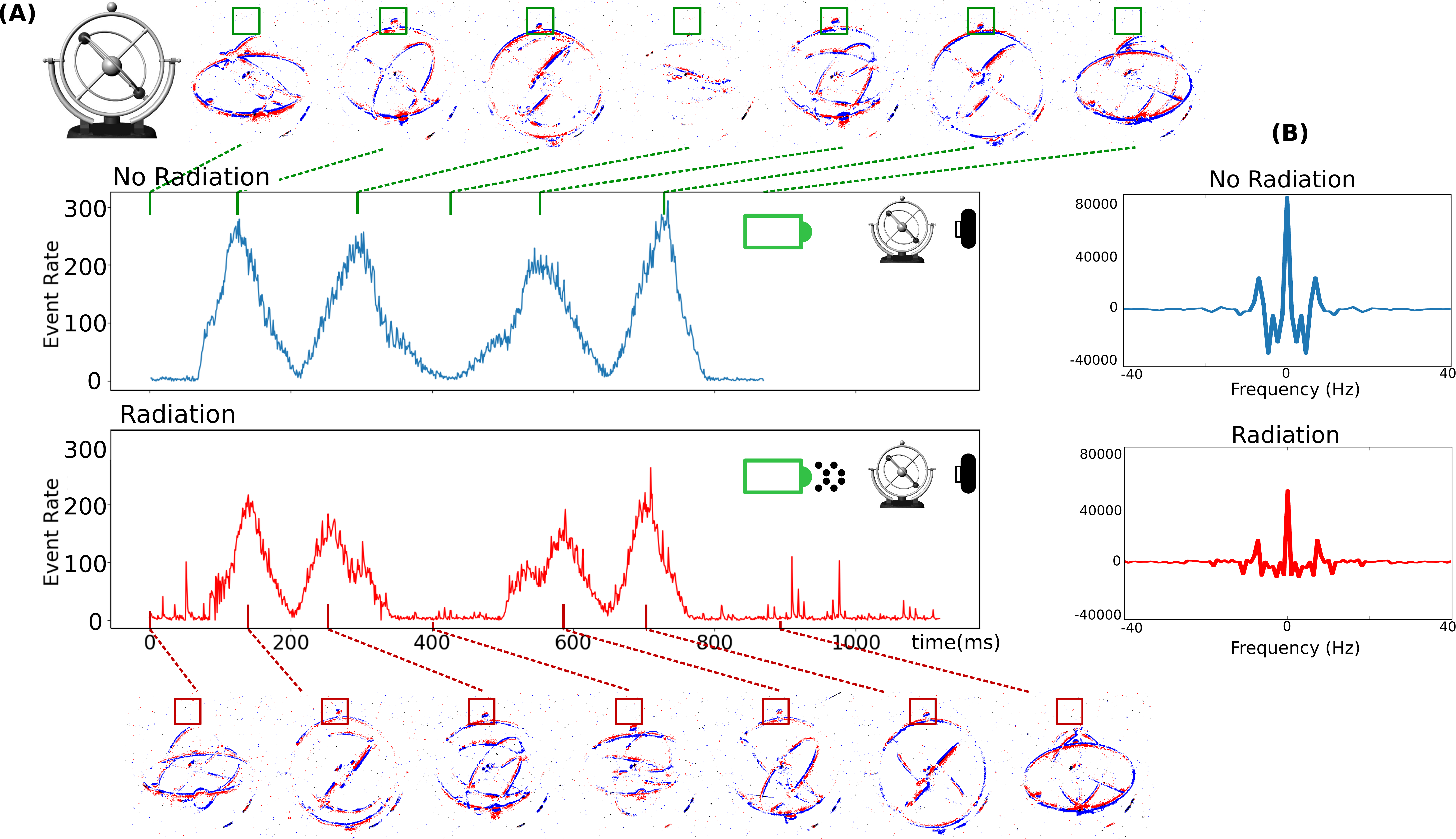}
    \caption{(A) An orbital pendulum recorded using the sensor and the event rate calculated as the number of events within a 1 ms moving window with (red) and without (blue) radiation turned on within a bounded box, as shown in the image panels. The images show the event frames obtained within the time window at different time points in the recording. Qualitatively, the sensor produced similar images for both conditions. (B) Calculated frequency of the pendulum using the event rates. The frequency of the pendulum's motion could be obtained using the FFT in each case.}
    \label{fig:withWithoutRad}
\end{figure*}

\subsection{Effects of Room Brightness} 
When deployed in space, these vision sensors may be subject to varying levels of background light intensity. To understand how neutron radiation would affect the sensor under such varying conditions, we recorded background noise events during radiation while placing the sensor in an artificially lit room with illuminance levels of around 500 \textit{lux} and with a lens cap covering the sensor to simulate a low-light intensity condition with a light level close to 0 \textit{lux}. The intrinsic characteristics of the sensor pixels allow them to be invariant to the background lighting conditions thanks to the relative change operation mode and the log scale. Figure \ref{fig:lightvsdark} shows the number of ON and OFF events induced by neutron radiation in the artificial lit "light room" and in the low-intensity "dark room" case. We find that the number of ON events induced in the dark room was nearly $1.5$ times higher than in the light room. Conversely, no significant difference were observed in the OFF events induced in the two conditions. Details of this process are explained in Section~\ref{sec:circuits}.

\subsection{Signal-to-Noise} 
In order to measure the signal-to-noise ratio, events were compared with the beam ON and OFF while the sensor observed a dynamic scene composed of a cyclic-pendulum, as shown in Figure \ref{fig:withWithoutRad}(A). 
To calculate the ratio between the two values, the number of signal events measured with the cyclic-pendulum were compared directly with the number of isolated radiation-induced noise events. This comparison can be seen in Figure \ref{fig:snr}. Comparing these values gives a signal-to-noise ratio of $3.355$.\par

To ensure the signal can be seen even when radiation is introduced, events in a $50\times50$-pixels bounding box (shown in green and red in Figure \ref{fig:withWithoutRad}(A)) were measured and plotted to compare signal data with and without radiation. Since the pendulum's movement is cyclic, we calculated the event rates data over time using a moving window of 1 ms. The frequency of this rate data calculated using Fourier transform should ideally give us the frequency of oscillation of the pendulum. The Fourier transform of the signal with and without radiation is shown in Figure \ref{fig:withWithoutRad}(B). With the addition of radiation noise, the signal's major frequency can still be estimated with some slight noise at low frequencies.

\begin{figure}[htbp]
    \centering
    \includegraphics[width=0.5\textwidth]{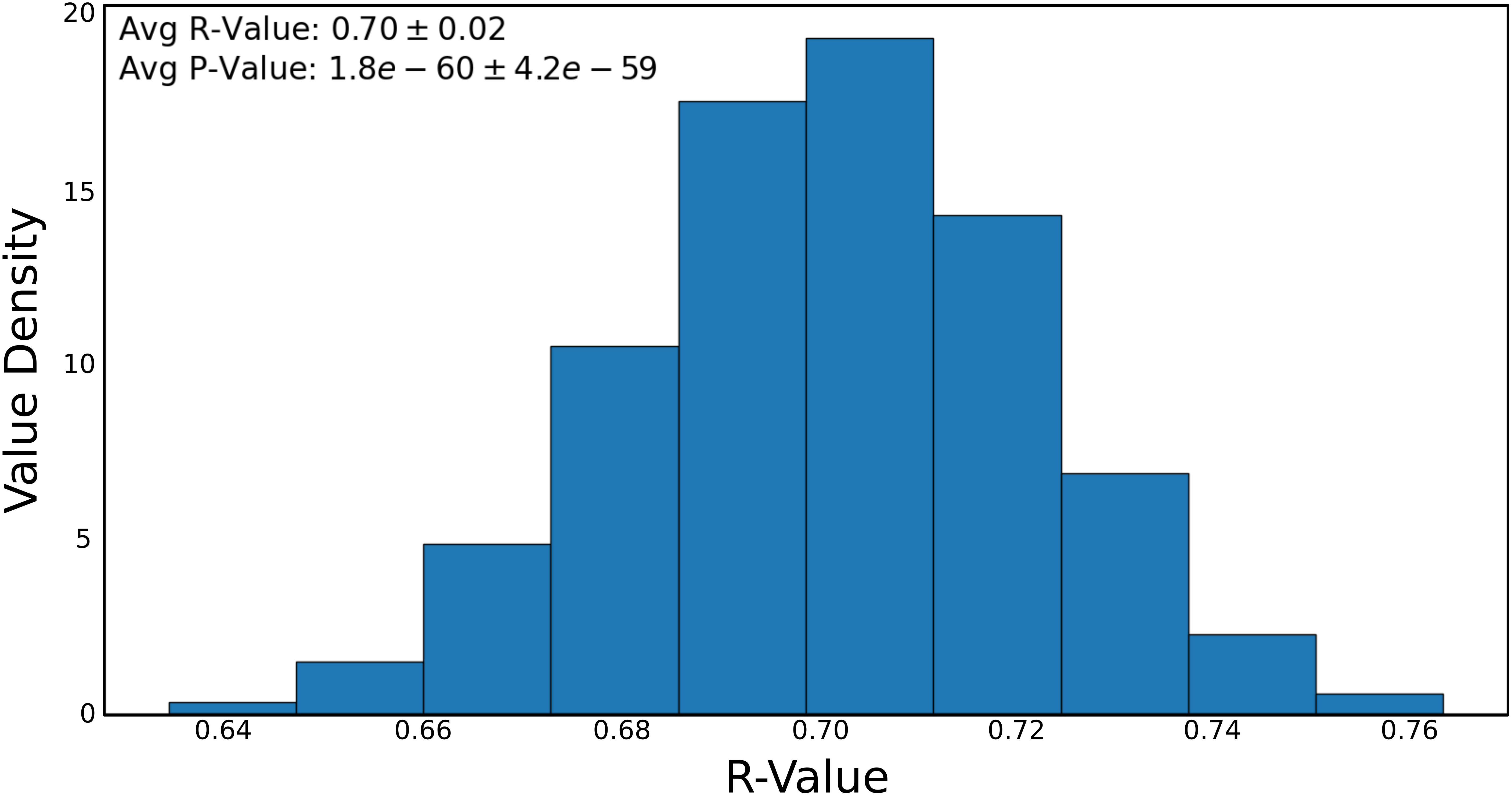}
    \caption{A Pearson correlation test was performed for the events obtained from the pendulum's movement with and without radiation. The high correlation and small standard deviation show that the signals obtained from the two conditions were quantitatively similar.}
    \label{fig:correlations}
\end{figure}

\begin{figure*}[htbp]
    \centering
    \includegraphics[width=0.9\textwidth]{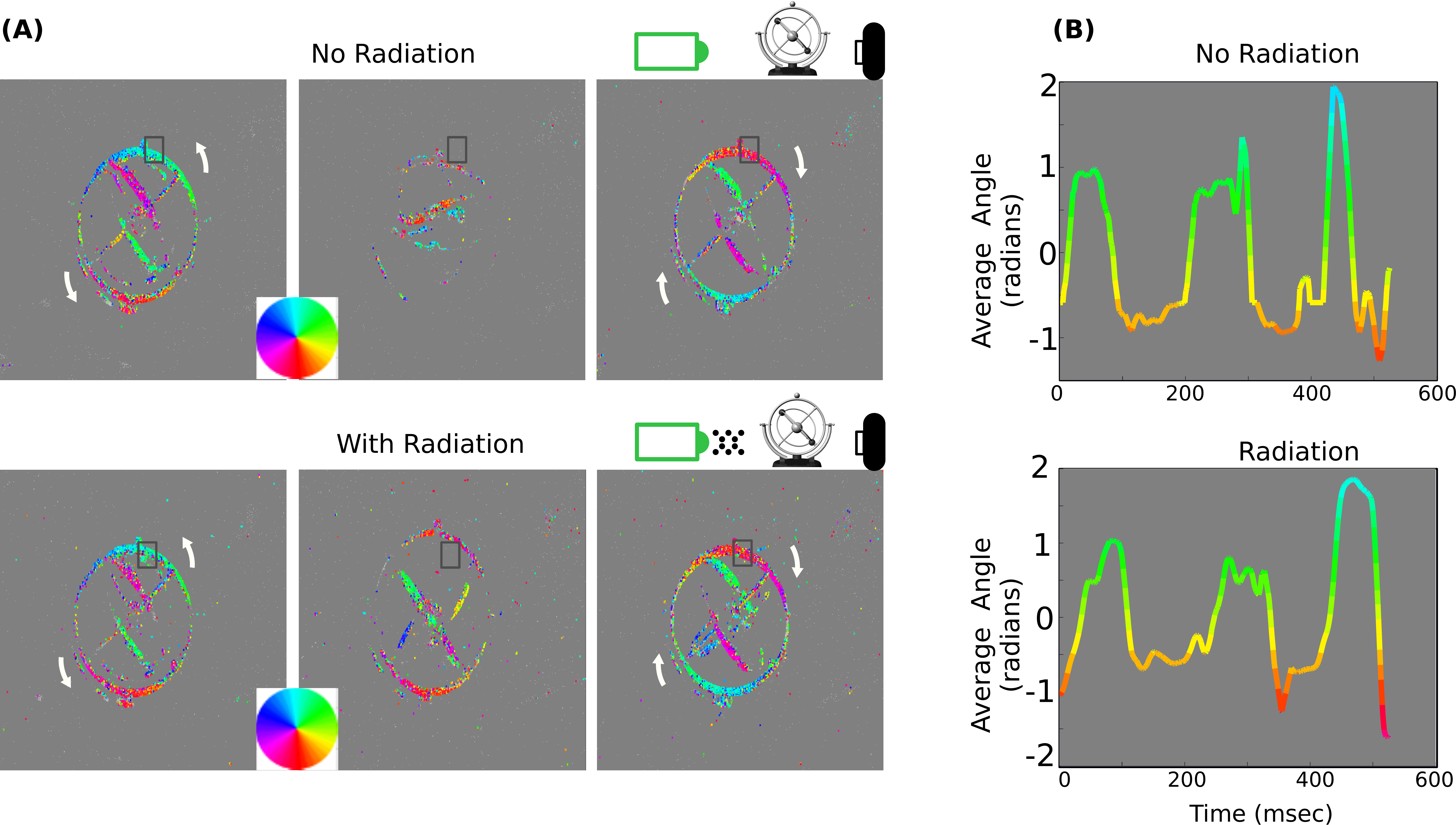}
    
    \caption{The movement directions of different parts of the pendulum system computed from the recorded event streams with and without radiation. (A) The colors represent the movement directions of the events as indicated by the color wheel. (B) Graphs showing the computed average movement directions for events occurring in a 5ms moving window within the black bounding box shown in the images. The Pearson correlation coefficient between two signals was 0.7189 indicating that the direction computation was not affected by the radiation.}
    \label{fig:opticalFlow}
\end{figure*}



To validate the signal-noise ratio of the radiated sensor, a Pearson-correlation test was performed between the radiation data and the non-radiation data. With a high correlation, it can be shown that the two distributions follow each other closely with minor linear transformations. Due to the varying size of samples, sub-samples were taken and analyzed to estimate the correlation R-value. The distribution of R-values can be seen in Figure \ref{fig:correlations}. The measured R-value was $0.70 \pm 0.02$ with a negligible p-value. It can therefore be deduced with high confidence that the radiation-induced noise is not enough to significantly change the data output from the original, non-radiated data.

The ultimate goal of deploying sensors on missions is to obtain useful information from them while in space. One of the most fundamental, low-level features that can be extracted from the event stream is motion flow. The optical flow provides the speed and direction of an object's movement in the camera plane, where its precision is related to the temporal properties of events. We computed optical flow on events captured from the sensor recording the moving pendulum system using the aperture-robust event-per-event optical flow technique introduced in \cite{akolkar2018realtime}. The average direction of movement of one arm of the pendulum inside a bounding box (shown in Figure~\ref{fig:opticalFlow}(A)) is plotted in Figure~\ref{fig:opticalFlow}(B). The average angle values follow the expected wave as the arm of the pendulum moves up and down vertically. The Pearson correlation between the two conditions was found to be 0.7189, showing that movement computation is not affected by radiation.

\begin{figure*}[h!]
    \centering
    \includegraphics[width=0.41\textwidth]{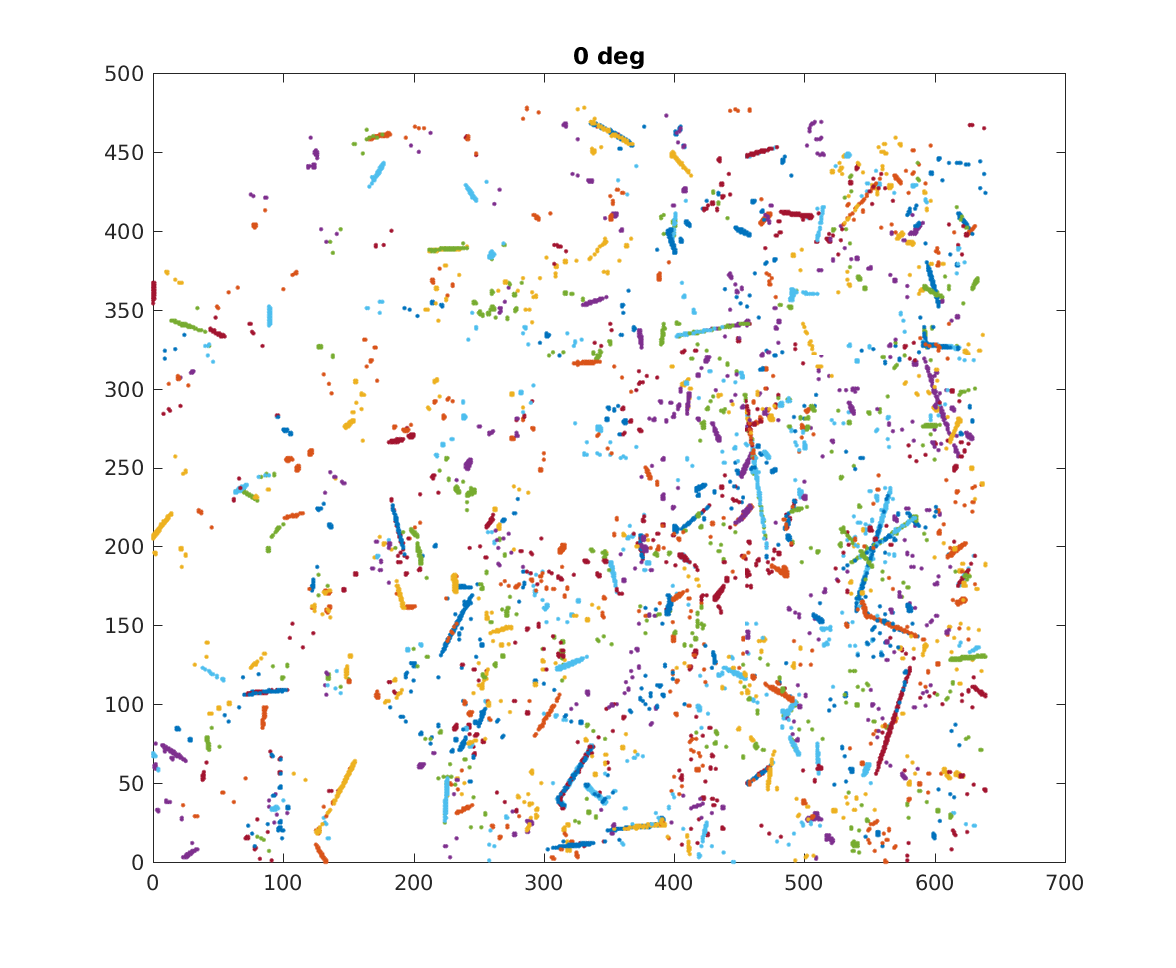}
    \includegraphics[width=0.41\textwidth]{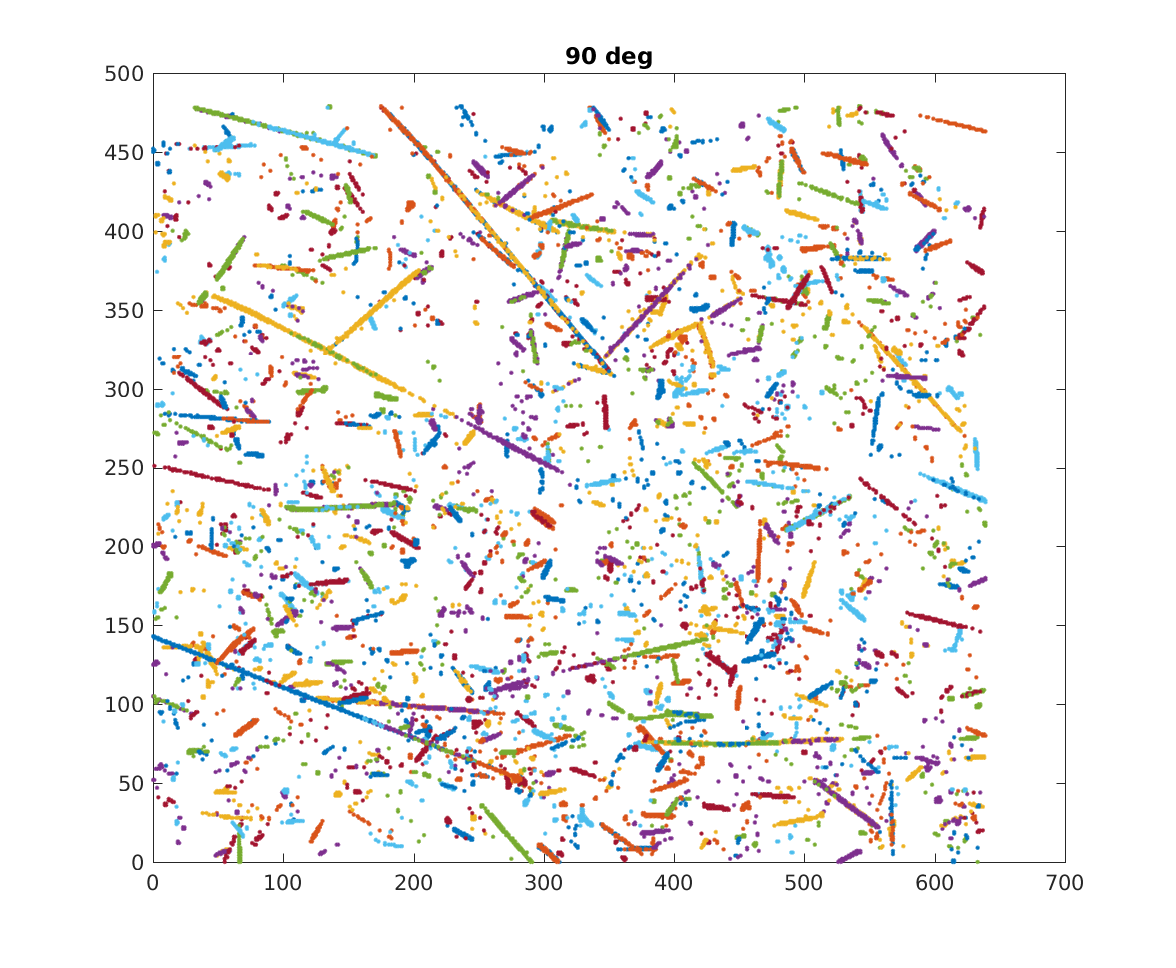}
    \caption{Clustered patterns of noise obtained by searching for clusters with minimum sizes of 10 pixels in 30-second recordings. Time slices of 5 ms were processed consecutively, searching for 10-pixel clusters. All clusters detected during 30 seconds are grouped in the plots. Events were recorded for a $0\degree$ angle of incidence and a $90\degree$ angle of incidence. Significantly more line segments can be seen at $90\degree$. At $0\degree$, fewer, smaller (average 30 pixels) and more clustered noise patterns were observed than at $90\degree$, where longer (up to 300 pixels) and more frequent (up to 7 times) line segments were observed.}
    \label{fig:patterns}
\end{figure*}

\begin{figure*}[h]
    \centering
    \includegraphics[width=0.9\textwidth]{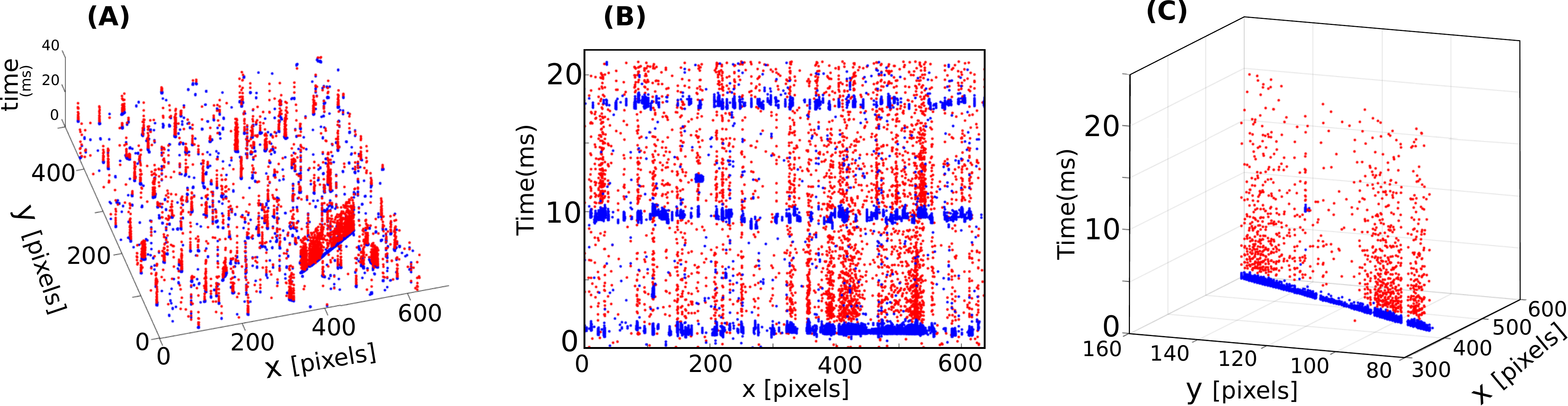}
    \caption{Details of 20ms event capture when exposed to neutron beam without visual stimulus. The blue dots represent positive events and the red dots represent negative events. Positive events are mostly concentrated in 600-800$\mu s$ time intervals separated by about 8 $ms$ intervals in which mostly negative events are recorded. (A) 3D plot (x,y,time) of events captured during the 20 $ms$ interval. Small scattered dots/clusters can be observed plus a line segment in the lower right part. (B) Time vs x-coordinate projection of the recorded events. (C) Events corresponding to the line segment in (A) which have been isolated for better visibility. }
    \label{fig:see_segs}
\end{figure*}

\begin{figure*}[h!]
    \centering
    \includegraphics[width=0.8\textwidth]{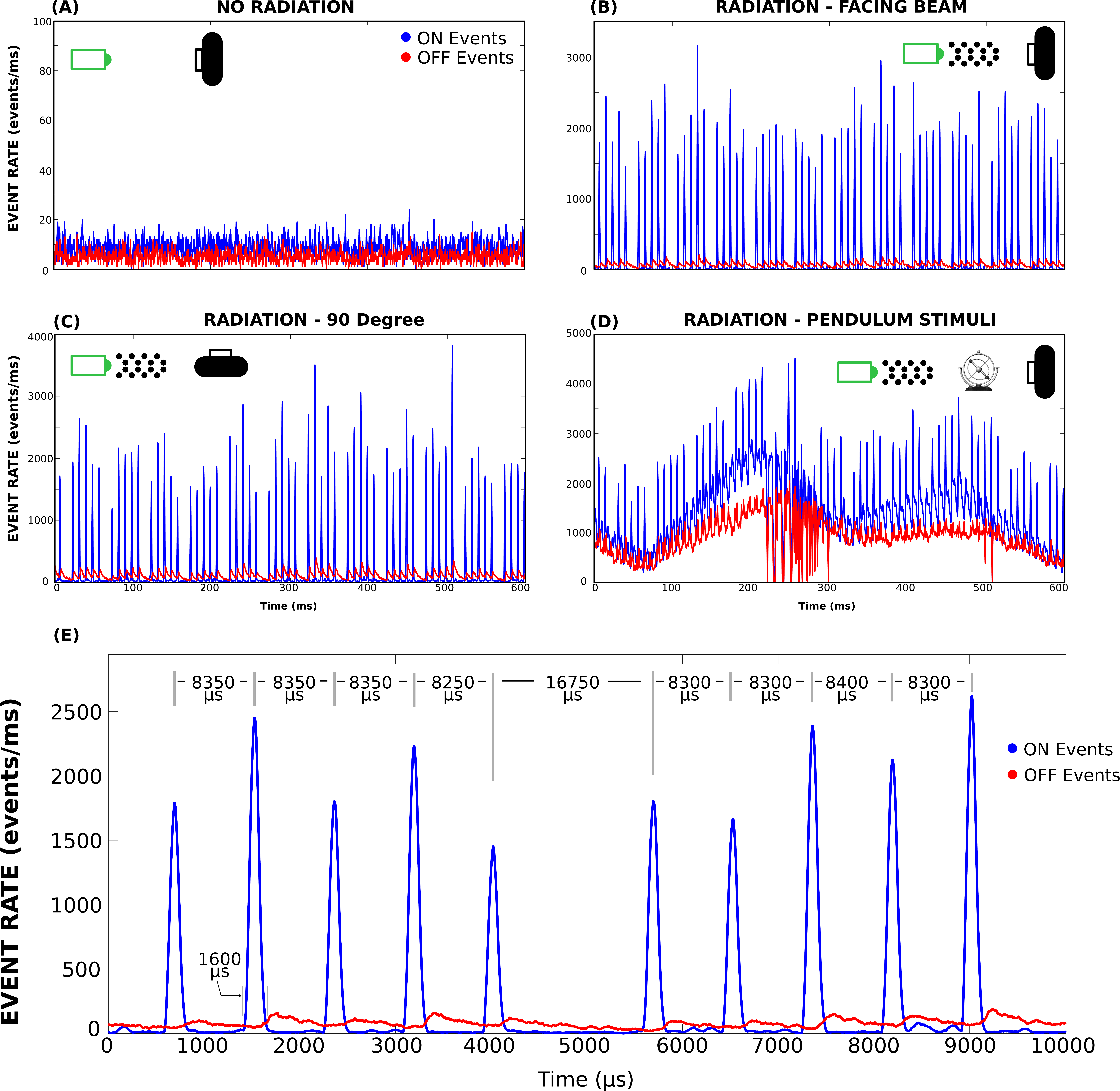}
    \caption{Noise rates for different conditions with and without neutron radiation. (A) The overall noise without radiation is very low. (B, C) Radiation noise when the sensor was placed at $0\degree$ (facing) (B) and at $90\degree$ (C) to the beam source. In each case, we recorded the bursts of noise most likely due to neutron pulses from beam generation. (D) Similar noise was found in the recording when a circular pendulum was recorded with the camera. The burst noise was superimposed on the low frequency events generated by the pendulum motion. (E) Details of the neutron's macro-pulse sequence can be observed from a zoomed-in plot of the event bursts in (B). Each neutron macro-pulse produced positive event bursts with duration of about 1.6 $ms$, and with peaks separated on average by 8.3 $ms$. Five macro-pulse responses appear, with a duration between the first and the fifth of 33.25 $ms$, while the time between two 5-macro-pulse trains is 16.75 $ms$.}
    \label{fig:noiseRates}
\end{figure*}

\subsection{Noise Patterns}
Radiation-induced noise, as shown in Figure \ref{fig:patterns} for both orientations, can be categorized into two main groups: clusters and line segments. Line segments represent a line of events that appear across the frame due to a neutron impacting the sensor at a non-zero angle of incidence. Clusters represent a random burst of events in a small area. The angle of incidence between the sensor and the radiation source affects the number of line segments. About 5-7 times more line segments appear with a $90\degree$ angle of incidence than with a $0\degree$ angle of incidence. Conversely, about twice as many clusters appear with a $0\degree$ angle of incidence than with a $90\degree$ angle of incidence. Significantly longer lengths of line segments occurred at $90\degree$, where streaks of up to 300 pixel lengths were observed, whereas smaller streaks, with maximum lengths of 30-50 pixels, were seen at $0\degree$. An example of differences in noise cluster patterns can be seen in Figure~\ref{fig:patterns}. These figures were obtained by analyzing recordings of $10^{7}$ events, with a duration about a 30-second, while searching for unconnected clusters not exceeding 10-pixel in size. Note that in Figure~\ref{fig:patterns}(A) more event density can be seen in the corner of high $x$ and low $y$ values than in the opposite corner. This is similar to what was observed in Figure~\ref{fig:density0deg} due to human error in placing the sensor in the beam path.\par

Analysis of noise line segments showed a burst of ON events over a fast time frame, followed by a long relaxation-period of OFF events after a short wait time, as shown in Figure~\ref{fig:see_segs}. This is due to an influx of positive current in the sensor's photo-diodes creating a burst of ON events, followed by a relaxation period for the current to return to normal, creating OFF events. The ON events burst over about 600-800$\mu s$ and the negative event tail is about 10ms long.\par

Viewing the event rate of the bursts, we see peaks of ON events followed by a long tail of OFF events. This effect is seen within all noise-types and is shown in Figures~\ref{fig:noiseRates} (B) and (C). Figure~\ref{fig:noiseRates}(E) shows a zoomed view with finer details. Bursts of 5 peaks separated by a time of 16.75 $ms$ can be seen. Each peak has a duration of about 1.6 $ms$ of positive events. Consecutive bursts are separated by 8.25 $ms$ within the 5 peaks. Consequently, on average, the five peaks occur every 33.25 $ms$ + 16.75 $ms$ = 50 $ms$, which is equivalent to 100Hz peaks. This coincides with the LANSCE neutron source description \cite{wender2020alamos}, where the neutron source emits a pulse of neutrons at a rate of about 100 Hz. Each such neutron peak is referred to as a neutron ``macro-pulse''.


\section{Circuit-Level Interaction Interpretation}
\label{sec:circuits}

High-energy neutron beams are thought of as ionizing radiation, which can instantaneously change the charge of an electric circuit node within the camera sensor chip. Since an event-camera can capture internal changes with microsecond resolution, these sensors provide a new way of ``seeing" fine interactions taking place between fast radiation particles and the electronic chip while it is operating.\par

\begin{figure}[H]
    \centering
    \includegraphics[width=\columnwidth]{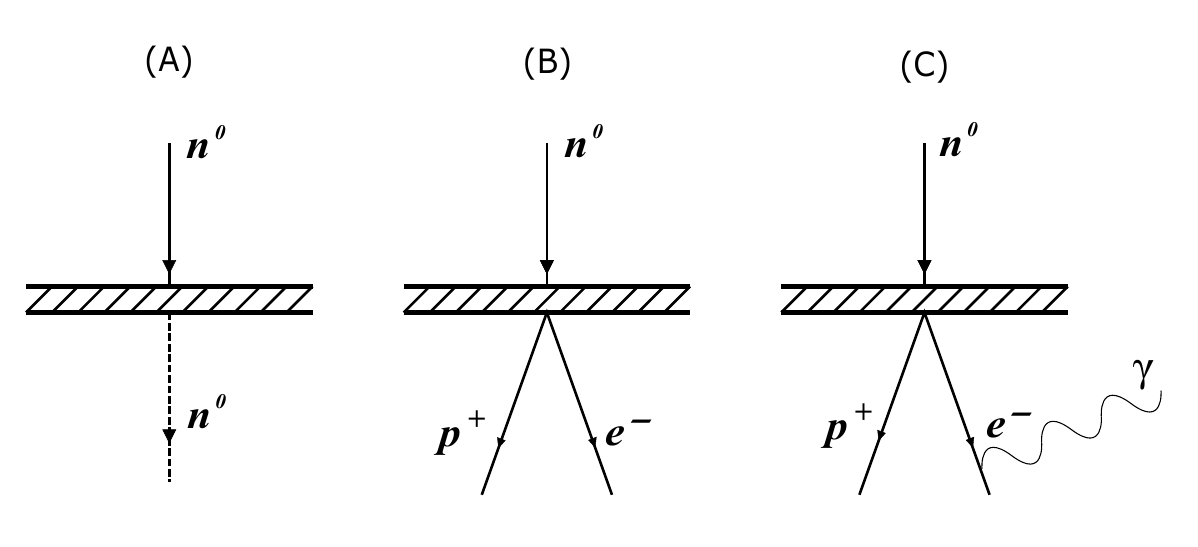}
    \caption{Three possible free neutron decays. (A) The neutron passes through the sensor casing without decaying. (B) The neutron decays into a proton and electron. (C) The neutron decays into a proton and an electron which emits gamma radiation.}
    \label{fig:neutron_diagram}
\end{figure}

For the free neutrons passing through the sensor, there are three main possibilities: the neutron can pass through as a neutron without decaying, the neutron can decay into a proton and an electron, or the neutron can decay into a proton and an electron which emits a gamma photon due to internal brehmsstrahlung \cite{particle2002review}. A diagram of these three possibilities can be seen in Figure \ref{fig:neutron_diagram}. Due to quantum uncertainties and the inability to distinguish between particles, it is impossible to distinguish the cases' impact on the sensor in this experiment. Further research must therefore be performed to detail the exact cause of the induced noise patterns.

In digital circuits, high-energy charged particles and radiation beams tend to mainly impact memory circuits, where charge is stored on tiny parasitic capacitors, producing bit-flips and consequently altering system states and data. In our sensor, however, we observed consistent sudden positive events over many pixels followed by negative event tails, 
synchronously with the macro-pulse neutron emission patterns of LANSCE \cite{wender2020alamos}. The fact that most responsive pixels produce a burst of positive events during each 625$\mu s$ LANSCE neutron macro-pulse, rules out the possibility that the sensor is suffering bit-flip effects at temporary memory-storing nodes. If this were the case, we would expect to observe a random mix of positive and negative events within each neutron macro-pulse. However, most of the affected pixels respond by providing a synchronized burst of positive events. It can thus be inferred that it is the pixels' photo-diodes that are responding to the neutron macro-pulses. Photo-diodes drive a photo-current proportional to incident light intensity. If a high-energy proton or electron crosses the depletion region of a photo-diode, it will interact, either by attraction or repulsion, with the electrons flowing through it at that moment, thus producing a sudden decrease in photo-current and, consequently, negative events. However, since we observed a sudden, very significant increase in photo-current (resulting in positive events), we hypothesize that the scattered pixels are sensing sudden radiation at their locations. This would also explain the observation of segments sensed simultaneously by consecutive pixels. Figure~\ref{fig:see_segs} shows one such segment in a 20 $ms$ time slice of events, corresponding to three consecutive 625$\mu s$ neutron macro-pulses separated from each other by 8.25 $ms$. Most of the pixel responses show small clusters of less than 10-pixels, the exception being the 190-pixel long segment. Our hypothesis is that the sensor is crossed by radiation bursts, most of them perpendicular to the chip plane, but occasionally interacting with deflected radiations at other angles and producing line segments. However, all radiation interactions occur precisely during the beam's macro-pulse times.

\begin{figure*}[]
    \centering
    \includegraphics[width=\textwidth]{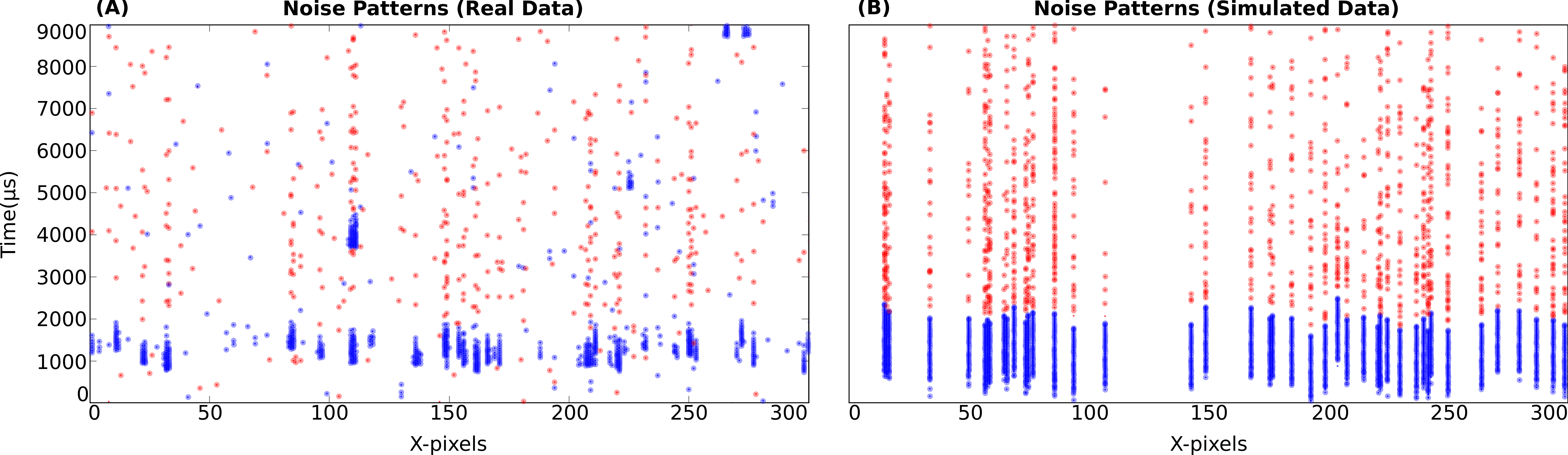}
    \caption{Examples of X-projections of noise line segments for events recorded from (A) real data and (B) simulated noise data. All cases show a burst of ON (blue) events with a long OFF (red) tail. As each example line segment was detected, the events were separated from the recording and then projected on the X-axis.}
    \label{fig:noise_real}
\end{figure*}

The electronic pixel circuitry of an event-camera chip has a limited response time in the range of 0.1 $ms$ to 10 $ms$ depending on ambient light and bias conditions~\cite{posch2011qvga}\cite{serrano2013}. The LANSCE neutron source macro-pulses have a time duration of 625 $\mu s$, which is lower than the temporal resolution of the event sensor. The macro-pulse radiation impinging on the destination pixels produces a sudden over-stimulation of photo-current, resulting in the sudden generation of a handful of positive events per pixel during the neutron macro-pulse. After such strong over-stimulation, the pixel circuit relaxes to its steady ambient-light-driven state with a time constant in the range of 10 $ms$, producing events of negative polarity over time. This behavior of sudden positive stimulation of 600-800 $\mu s$, where positive events are produced, followed by about 8-10 $ms$ of negative-event relaxation is systematically observed in the recordings. Figure~\ref{fig:see_segs}(A) shows the 20 $ms$ event capture with scattered noise-like dots/clusters of fast positive events (shown in blue), followed by negative event tails (shown in red). We hypothesize that each such dot/cluster corresponds to a neutron crossing the chip. Figure~\ref{fig:see_segs}(B) shows the events in Figure~\ref{fig:see_segs}(A), but displayed in their corresponding time vs x-coordinate projection. We can clearly see the synchronized sequence of neutron macro-pulse-induced positive events (shown in blue), of 600-800 $\mu s$ duration, separated by about 8 $ms$ of inter-neutron macro-pulse time where mainly negative relaxation events are produced. The figure also shows a 190-pixel long segment with the same time profile. The events for this segment are isolated in Figure~\ref{fig:see_segs}(C). In this plot there are 2,031 positive events collected over about 800 $\mu s$, followed by 1,090 negative events collected during over about 20 $ms$. \par

The suddenly induced photo-current hypothesis also explains the observations in Figure~\ref{fig:lightvsdark}, where more positive events are produced under dark-room conditions than under light-room conditions. When under light room conditions, the photo-diodes are already driving some current and consequently reach their maximum saturation current earlier when suddenly impinged by high energy particles, resulting in fewer induced positive events. Under dark conditions, the photo-current can undergo a larger variation, resulting in more positive events.

\section{Event-RINSE Simulator}

\begin{figure}[h]
    \centering
    \includegraphics[width=.85\columnwidth]{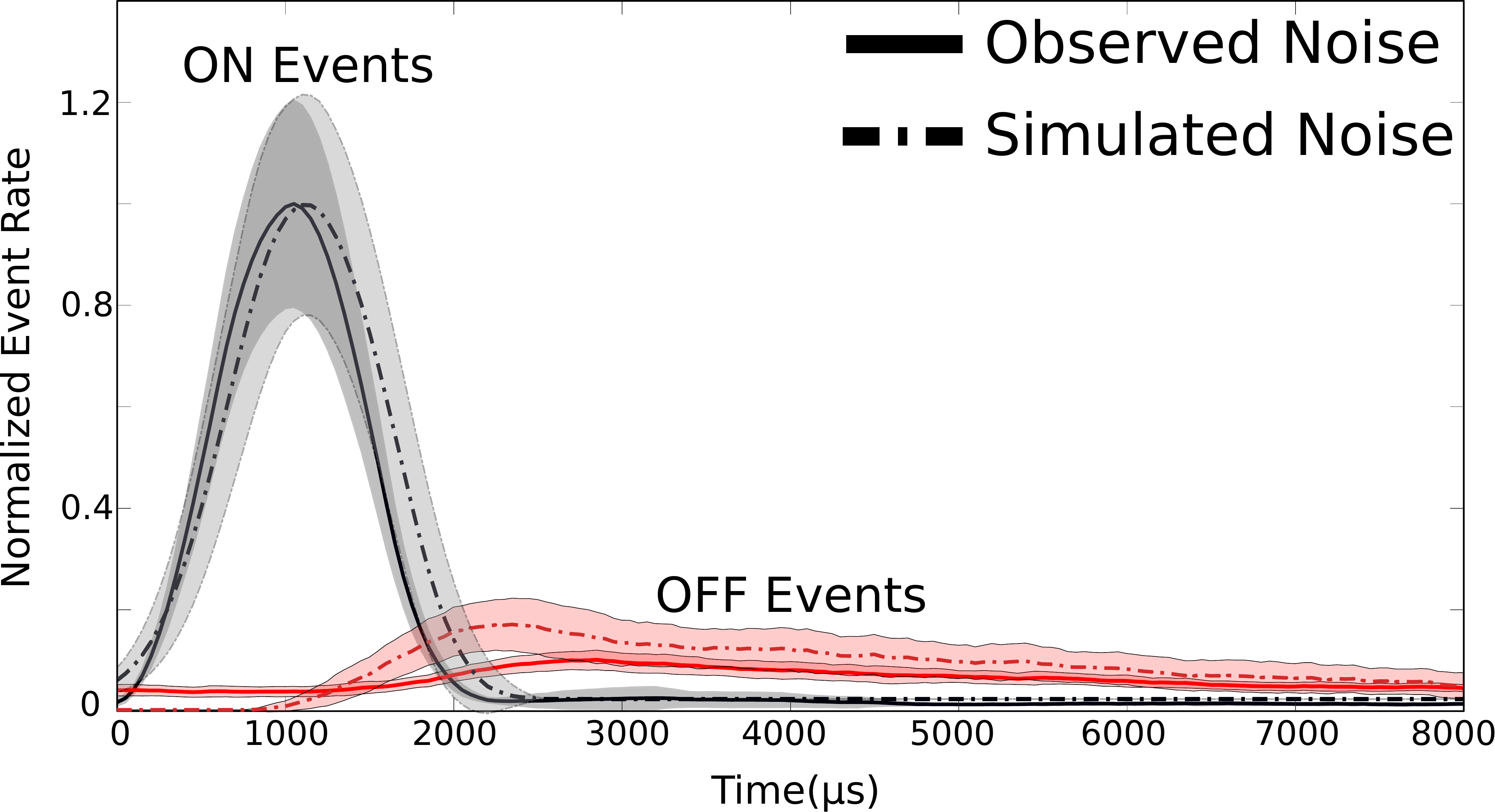}
    \caption{Average single pixel radiation-induced event rate model for observed and simulated data. From real data we observed that neutron interactions induced ON event bursts of about 1.6 $ms$ within the first 1 $ms$. These were followed by long tails of OFF events lasting up to 10 $ms$. The simulator was used to induce noise events into the stream of recorded non-noisy data, and the noise characteristics for single-event noise were then averaged to create the dashed curves. The simulator was able to match the real noise model within a margin of acceptable error.}
    \label{fig:pattern_validation}
\end{figure}


\subsection{Simulated Noise Generation}

Given a stream of event-camera data as input, the simulator steps through each event. For every time step in the data, a noise event is either generated or passed. The probability of injection was determined using a Poisson distribution of observing $k=1$ event with a variable event rate. Namely, 
\begin{equation}
P(\lambda) = \lambda e^{-\lambda}
\label{eqn:poisson}
\end{equation}
\noindent where $\lambda$ is the frequency of an event happening per microsecond. A starting pixel is randomly chosen uniformly across the resolution of the sensor. The simulator decides whether injected noise is in the form of a cluster or a line segment based on the angle-of-incidence parameter. Specifically, the chance of injecting a cluster is based on the cosine of the angle of incidence with some jitter-error. Thus, the probability of the injected noise pattern is given by Eq~\ref{eqn:patternProb}.

\begin{equation}
\begin{gathered}
P(Cluster) = |cos(\theta + \varepsilon))| \\
P(Line\ Segment) = 1 - P(Cluster)
\end{gathered}
\label{eqn:patternProb}
\end{equation}

\noindent where $\theta$ is the angle of incidence in radians and $\varepsilon$ is a small amount of error. A cluster's shape is modelled by randomly chosen pixels around the neighborhood of the starting point. A line segment is modelled by a straight line with a randomly chosen angle between 1$\degree$ and 360$\degree$.\par

For each pixel in the shape of the generated event, the noise pattern is modelled by sampling a time window for ON events from $\mathcal{N}(2000 \mu s,200\mu s)$ which represents the length of time for the burst of ON events. OFF events are sampled from $\mathcal{N}(8000 \mu s,1000 \mu s)$. More precisely,
\begin{equation}
\begin{gathered}
        P(ON\ Noise\ Event; t) = e^{-\frac{1}{2}\frac{(t-t_{ON})^{2}}{\sigma_{ON}}} \\
        t \in [0,~t_{ON}] \\
        t_{ON} \sim \mathcal{N}(2000,200)
\end{gathered}
\label{eqn:ONgeneration}
\end{equation}

\noindent where the burst of ON events is simulated as a Gaussian model with the mean as the sampled ON event time window ($t_{ON}$) and standard deviation $\sigma_{ON}$ $=$ $340$~$\mu s$ is used to determine the probability of generating an event over time $t$. 
The wait time between the burst of ON events and the OFF-event relaxation period is sampled from $\mathcal{N}(100,50)$. After the wait time ($t_{Wait}$), the current relaxation of OFF events is modelled using an exponential with decay parameter, $\beta$ $=$ 5200 up to a total OFF time ($t_{OFF}$) as per Eq.(\ref{eqn:OFFgeneration}).

\begin{equation}
    \begin{gathered}
    P(OFF\ Noise\ Event; t) = \frac{1}{\beta}e^{-\frac{1}{\beta}t} \\
    t \in [t_{ON} + t_{Wait},~t_{OFF}] \\
    t_{Wait} \sim \mathcal{N}(100,50)\\ 
    t_{OFF} \sim \mathcal{N}(8000,1000)
    \end{gathered}
    \label{eqn:OFFgeneration}
\end{equation}

The generated events are then added to the data file and sorted by timestamp in ascending order. Finally, the file is saved to be used in testing or evaluation. The algorithm to generate radiation-induced noise events is detailed in Algorithm \ref{algorithm:main}.

\subsection{Pattern Validation}
To validate the simulation environment, noise events were generated following the pattern described in Algorithm \ref{algorithm:main} and compared with noise events from real data. The noise events were plotted against time to compare them with noise from observations. Figure \ref{fig:noise_real} shows a sample of visual real noise events (Figure \ref{fig:noise_real}(A)) vs simulated noise events (Figure \ref{fig:noise_real}(B)). The model used to generate noise was compared to the average observed single-event noise. The model shown in Figure \ref{fig:pattern_validation} fits the observed pattern with a $5\%$ error  rate for ON noise profiles and $12.3\%$ error rate for OFF noise profile. 

\begin{algorithm}[ht]
\begin{algorithmic}[1]
\caption{Radiation Induced Noise Simulation Environment (Event-RINSE)}
\For{At each time step $t$} \label{Each_time_step}
	\State{Compute chance of radiation-induced noise using Eq.(\ref{eqn:poisson})}
    \If{Generate Noise Event}
        \State{Decide if noise is cluster or line using Eq.(\ref{eqn:patternProb})}
        \State{Choose a random pixel $[x_0, y_0]$}
            \If{CLUSTER NOISE}
                \State{Randomly sample a set of pixels {$[X, Y]$} in the neighborhood of $[x_0, y_0]$}
                \For{Each pixel $\in$ {$[X, Y]$} in the cluster}
                        \State{Generate ON Events Using Eq.(\ref{eqn:ONgeneration})}
                        \State{Generate OFF Events Using Eq.(\ref{eqn:OFFgeneration})}
                \EndFor
            \EndIf
            \If{LINE NOISE}
                \State{Randomly sample angle of line: $\theta$ $\in$ [0, 2$\pi$)}
                \State{Select a set of pixels {$[X, Y]$} forming a line L starting at $[x_0, y_0]$ with angle $\theta$}
                    \For{Each pixel {$[X,Y]$} of the line}
                        \State{Generate ON Events Using Eq.(\ref{eqn:ONgeneration})}
                        \State{Generate OFF Events Using Eq.(\ref{eqn:OFFgeneration})}
                    \EndFor
            \EndIf
    \EndIf
    \State{Append noise events to stream}
\EndFor
\State{Sort events by ascending timestamps}
\label{algorithm:main}
\end{algorithmic}
\end{algorithm}

\subsection{Simulation Environment Usage}

The Event-RINSE simulation environment is written in Python with many supporting parameter flags that can be used to modify the simulation model. Normal Python data analysis modules are needed for the simulator, namely SciPy \cite{2020SciPyNMeth} and NumPy \cite{oliphant2006guide}, while OpenCV \cite{opencv_library} is used to display videos of the event data. The simulator is run using Python3 environment with runtime flags for campaign customization. Currently available flags and descriptions can be seen in Table \ref{table:rinseOptions}. The input data file is the only input that is necessary to run the simulator. Input files are assumed to be plain text files in $<x>\ <y>\ <timestamp\ (\mu s)>\ <polarity>$ format.

\begin{table*}[ht]
    \centering
    \caption{Summary of Event-RINSE runtime options.}
    \begin{tabularx}{\textwidth}{||Y Y Y||} 
        \hline
         Command Flag & Description & Datatype \\ [0.5ex] 
         \hline\hline
         -h/--help & Display help message and exit & N/A \\ 
         \hline
         -f/--input-file & The input data file path to read from & String \\
         \hline
         -o/--output-file & Custom output data file path to write to & String \\
         \hline
         -aoi/--angle-of-incidence & Angle of incidence between the sensor and simulated beam. Affects prevalence of lines vs. clusters & Integer \\
         \hline
         -s/--imgSize & The size of the images from the sensor data & List of 2 integers \\ 
         \hline
         -vi/--view-input & View the input data file as a video & N/A \\
         \hline
         -vo/--view-output & View the output data file as a video & N/A \\
         \hline
         -i/--inject & Perform injections on input file and write to output file & N/A \\
         \hline
         -d/--delta & Time-step to hold in one frame when viewing video & Float \\
         \hline
         -n/--noise & The event rate of noise with standard deviation & List of 2 integers \\ 
         \hline
    \end{tabularx}
    \label{table:rinseOptions}
\end{table*}

\section{CONCLUSION}

The purpose of this experiment was to irradiate an event-based camera under wide-spectrum neutrons to view and classify any SEEs that may be observed. The results show that the main SEU that affects the event-based camera is radiation-induced noise in the form of uniformly-distributed events across the sensor's field of view. 
We found that noise induced on single pixels resulted in both ON and OFF events with a ratio of 3:1. An average noise event rate was found to generate peaks with lags in the range of 8-10 $ms$ which corresponded directly with the macro-pulse patterns of the neutron source at LANSCE \cite{wender2020alamos}. This shows that the sensor acted like a naive particle detector, and was only affected by the radiation over short timescales. 
OFF events were also seen to follow the ON-event peaks with exponentially-decaying event-rate profile. These profiles seem to suggest that the neutrons interact with the photo-diode in individual pixels causing energy dumps leading to large photo-current, inducing the ON events in a short time period of about 1.6 $ms$. The residual relaxation current after the radiation passes gives rise to the OFF events at much lower rates, but with a longer duration of up to 10 $ms$. The radiation did not cause any permanent, long-term damage to the sensor's photo-diodes or the hardware circuitry.
This hypothesis was further confirmed when looking at the noise events in brighter and darker background-illumination conditions, where ON events were significantly higher in the dark environment due to sensor's higher contrast sensitivity but OFF events were not found to change significantly across the two conditions. \par

Focusing on induced noise, experiments were performed to observe correlations with the angle of incidence and the event rate through the sensor. Surprisingly, the null hypothesis that there is no correlation between the number of events and the angle of incidence, was supported. With a larger angle of incidence, the cross-sectional area of the sensor is smaller to the beam's point-of-view, making it less likely to be hit. When a neutron does impact the sensor, however, it travels across the field leaving a long streak of events following its trajectory. When there is a smaller angle of incidence, the sensor looks larger from the perspective of the beam. This implies that the sensor will be more likely to be hit, but events are shown only in the form of dots as short lines of neutrons penetrate the sensor. These two effects thus cancel each other out, showing no difference in the induced event rate. \par

Comparing the number of events from a pendulum signal with radiation-induced noise shows a signal-to-noise ratio of 3.355. This ratio demonstrates the robustness of the event-based sensor to radiation in that the noise introduced does not significantly impact its ability to extract features of the desired signal. This is further illustrated by the sensor's ability to clearly observe the sinusoidal signal against the noisy background, and by the results of the optical flow algorithm implemented on the recorded events, which show no significant deterioration between the flow directions computed from the events when the radiation is introduced.\par


The Event-RINSE simulation environment created using the recorded noise data can be used to simulate the effects of radiation on pre-recorded data files. Event-RINSE was used to inject noise into the event streams recorded without radiation and was found to correspond well with the observed profile. The noise examples generated from the simulator matched both the average single-event noise model and the average noise across the sensor.
This fault injector makes it possible to test different neuromorphic-sensor algorithms, such as object tracking, under a noisy radiation environment without the need for expensive radiation testing, and thereby to assess an algorithm's viability in space and in any noise suppression techniques. Future work could look at improving the parameters and probability models for more accurate noise generation.

Further development of event-cameras for space should include research into their efficacy under proton and heavy-ion radiation. These experiments will show if the sensor, as it currently stands, is capable of survival under the harsh conditions of space. Future work could also include testing the sensor's capability to perform basic object tracking under neutron irradiation. The noise shown in this experiment could pose a small problem for SSA by interfering with signal events in object tracking. However, since the noise was seen to be fairly constant under various cases, it could be modeled for background analysis. Also, the induced noise did not appear to deteriorate signal analysis enough to cause detrimental effects. With minor background suppression, the signal-to-noise ratio could therefore be improved enough to perform the necessary algorithms and analysis for SSA on future spacecraft.

\section*{ACKNOWLEDGMENTS}

This research was supported by SHREC industry and agency members and by the IUCRC Program of the National Science Foundation  under Grant No. CNS-1738783. This work was performed, in part, at the Los Alamos Neutron Science Center (LANSCE), a NNSA User Facility operated for the U.S. Department of Energy (DOE) by Los Alamos National Laboratory (Contract 89233218CNA000001). We would also like to thank M. Lozano, M. Ull{\'a}n, S. Hidalgo, C. Fleta, and G. Pellegrini from the "Instituto de Microelectr{\'o}nica de Barcelona" (IMB-CSIC) for insightful discussions.

\newpage
\bibliographystyle{IEEEtran}
\bibliography{citation} 
\newpage

\newpage

\begin{IEEEbiography}[{\includegraphics[width=1in,height=1.25in,clip,keepaspectratio]{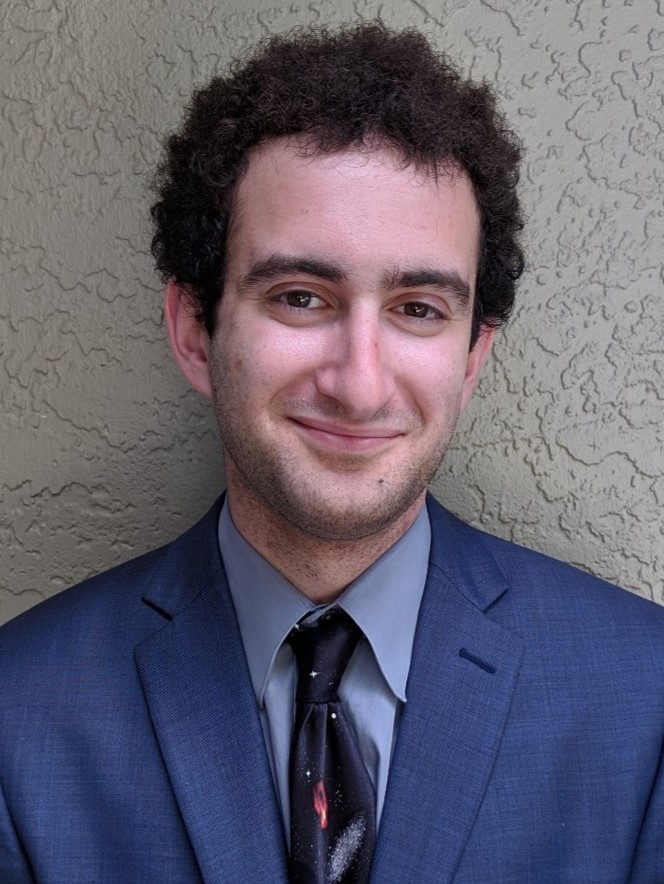}}]{Seth Roffe} earned a Bachelor of Philosophy in Physics, Astronomy, and Mathematics from the University of Pittsburgh in 2017 and a M.S degree in Electrical and Computer Engineering from the University of Pittsburgh in 2020. He is currently pursuing a PhD in Electrical and Computer Engineering. 

He has been a member of the NSF Center for Space, High-performance and Resilient Computing (SHREC) since 2018 performing research in space computing under the direction of Dr. Alan George. His main research interests involve resilience in sensor processing including data reliability and error classification in novel sensors.
\end{IEEEbiography}

\begin{IEEEbiography}[{\includegraphics[width=1in,height=1.25in,clip,keepaspectratio]{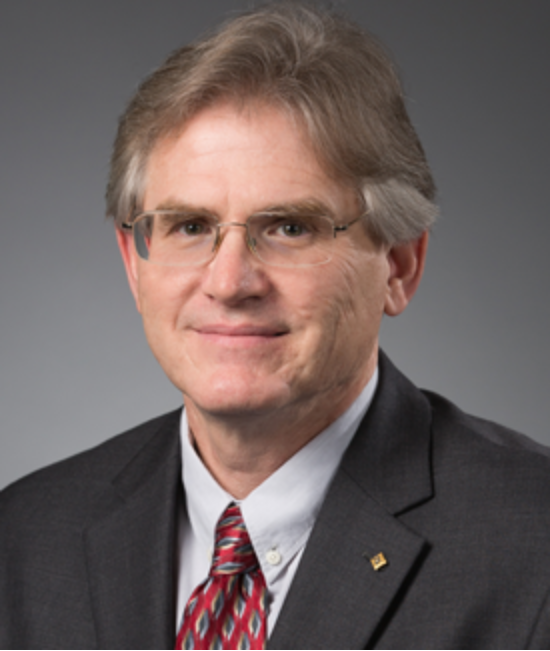}}]{Alan D. George} is Department Chair, R\&H Mickle Endowed Chair, and Professor of Electrical and Computer Engineering (ECE) at the University of Pittsburgh. He is Founder and Director of the NSF Center for Space, High-performance, and Resilient Computing (SHREC) headquartered at Pitt. SHREC is an industry/university cooperative research center (I/UCRC) featuring some 30 academic, industry, and government partners and is considered by many as the leading research center in its field. Dr. George's research interests focus upon high-performance architectures, applications, networks, services, systems, and missions for reconfigurable, parallel, distributed, and dependable computing, from spacecraft to supercomputers. He is a Fellow of the IEEE for contributions in reconfigurable and high-performance computing.
\end{IEEEbiography}

\begin{IEEEbiography}[{\includegraphics[width=1in,height=1.25in,clip,keepaspectratio]{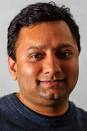}}]{Himanshu Akolkar} is  currently  a Post Doctoral Associate  at the University of Pittsburgh. He received his M.Tech. degree from IIT, Kanpur (India) in EE  and PhD from IIT, Genoa (Italy) in Robotics after which he had a Post Doctoral stint at Universite Pierre et Marie Curie. His primary interest is to understand the neural basis of sensory and motor control to develop an intelligent machine.
\end{IEEEbiography}

\begin{IEEEbiography}
[{\includegraphics[width=1in,height=1.25in,clip,keepaspectratio]{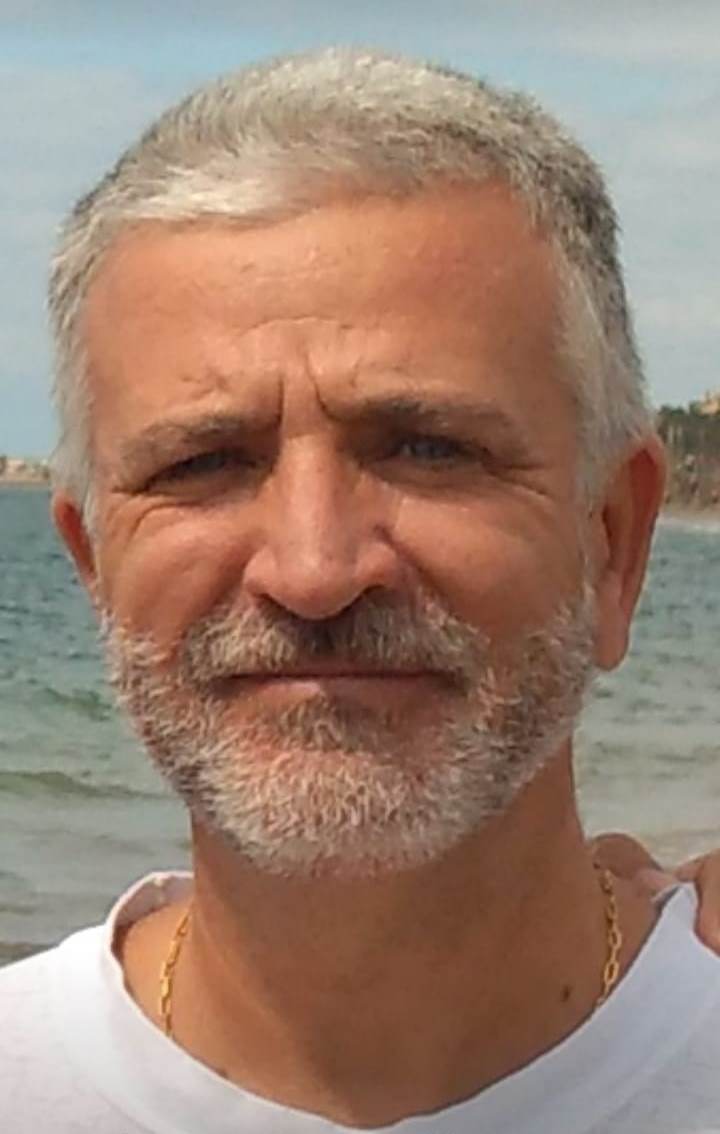}}]{Bernab{\'e} Linares-Barranco} (M'90-S'06-F'10) received the B.S. degree in electronic physics in June 1986 and the M.S. degree in microelectronics in September 1987, both from the University of Seville, Sevilla, Spain. From September 1988 until August 1991 he was a Graduate Student at the Dept. of Electrical Engineering of Texas A\&M University. He received a first Ph.D. degree in high-frequency OTA-C oscillator design in June 1990 from the University of Seville, Spain, and a second Ph.D. degree in analog neural network design in December 1991 from Texas A\&M University, College-Station, USA. Since June 1991, he has been a Tenured Scientist at the "Instituto de Microelectr{\'o}nica de Sevilla", IMSE-CNM (CSIC and Univ. de Sevilla), Sevilla, Spain. In January 2003 he was promoted to Tenured Researcher, and in January 2004 to Full Professor. Since February 2018, he is the Director of the "Insitituto de Microelectr{\'o}nica de Sevilla". He has been involved with circuit design for telecommunication circuits, VLSI emulators of biological neurons, VLSI neural based pattern recognition systems, hearing aids, precision circuit design for instrumentation equipment, VLSI transistor mismatch parameters characterization, and over the past 25 years has been deeply involved with neuromorphic spiking circuits and systems, with strong emphasis on vision and exploiting nanoscale memristive devices for learning. He is co-founder of two start-ups, Prophesee SA (www.prophesee.ai) and GrAI-Matter-Labs SAS (www.graimatterlabs.ai), both on neuromorphic hardware. He is an IEEE Fellow since January 2010. He is Chief Editor of Frontiers in Neuromorphic Engineering since 2021.
\end{IEEEbiography}

\begin{IEEEbiography}
[{\includegraphics[width=1in,height=1.25in,clip,keepaspectratio]{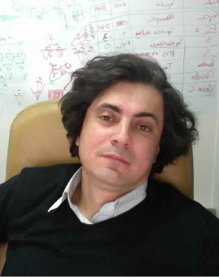}}]{Ryad   Benosman} received  the  M.Sc.  and  Ph.D. degrees  in  applied  mathematics  and  robotics  from University Pierre and Marie Curie in 1994 and 1999, respectively. He is Full Professor at the university of Pittsburgh/Carnegie Mellon/Sorbonne University. His work pioneered the field of event based vision.  He is the cofounder of several neuromorphic related companies including Prophesee and Pixium Vision a french prosthetics company. Ryad Benosman has authored more than 60 publications that are considered foundational to the field of event based vision and holds several patents in the  area of vision, robotics and image  sensing. In  2013, he was awarded with the national best French scientific paper by the publication LaRecherche for his work on neuromorphic retinas applied to retina prosthetics.
\end{IEEEbiography}

\EOD
\end{document}